\documentclass[useAMS,usenatbib]{mn2e}

\usepackage{enumerate}
\usepackage{amsmath,amssymb,amsfonts}
\usepackage{epsf,lscape}
\usepackage{colordvi}
\usepackage{color}
\usepackage[dvipdfmx]{graphicx}
\usepackage[colorlinks, linkcolor=blue, citecolor=black,dvipdfmx]{hyperref}

\def\<{\langle}\def\>{\rangle}
\def\v{\mathbf{v}}
\def\k{\mathbf{k}}\def\p{\mathbf{p}}
\def\s{\mathbf{s}}
\def\L{\mathcal{L}}

\def\dD{\delta_{\rm D}}

\title[Constraints on $f$, $D_{\rm A}$, $H$ with the SDSS LRGs]{
Simultaneous constraints on the growth of structure and cosmic expansion 
from the multipole power spectra of the SDSS DR7 LRG sample
}
\author[A. Oka et al.]{
\parbox{\textwidth}{Akira~Oka$^{1}$\thanks{E-mail: oka@utap.phys.s.u-tokyo.ac.jp}, 
Shun~Saito$^{2}$,
Takahiro~Nishimichi$^{3}$, 
Atsushi~Taruya$^{2,4,5}$,\\
Kazuhiro~Yamamoto$^{6}$}
\\
\\
$^{1}$ Department of Physics, University of Tokyo, Tokyo 113-0033, Japan\\
$^{2}$ Kavli Institute for the Physics and Mathematics of the Universe (WPI), University of Tokyo, Chiba 277-8583, Japan\\
$^{3}$ Institute d'Astrophysics de Paris, CNRS, 98 bis Boulevard Arago, F-75014 Paris, France\\
$^{4}$ Research Center for the Early Universe, School of Science, University of Tokyo, Tokyo 113-0033, Japan\\
$^{5}$ Yukawa Institute for Theoretical Physics, Kyoto University, Kyoto 606-8502, Japan\\
$^{6}$ Department of Physical Science, Hiroshima University, Higashi-Hiroshima 739-8526, Japan
}

\date{\today}
\begin{document}

\label{firstpage}

\maketitle

\begin{abstract}
The anisotropic galaxy clustering on large scales provides us with a unique opportunity to probe 
into the gravity theory through the redshift-space distortions (RSDs) and the Alcock-Paczynski effect. 
Using the multipole power spectra up to hexadecapole ($\ell=4$), of the Luminous Red Galaxy 
(LRG) sample in the data release 7 (DR7) of 
the Sloan Digital Sky Survey II (SDSS-II), we obtain simultaneous constraints on the linear growth rate
$f$, angular diameter distance $D_{\rm A}$,
and Hubble parameter $H$ at redshift $z = 0.3$. 
For this purpose, we first extensively examine the validity of a theoretical model for 
the non-linear RSDs using mock subhalo catalogues from $N$-body simulations, which are constructed 
to match with the observed multipole power spectra. 
We show that the input cosmological parameters of the simulations can be recovered well within 
the error bars by comparing the multipole power spectra of our theoretical model and those
of the mock subhalo catalogues.
We also carefully examine systematic uncertainties in our analysis by testing the dependence on 
prior assumption of the theoretical model and the range of wavenumbers to be used in the 
fitting.
These investigations validate that the theoretical model can be safely applied
to the real data.
Thus, our results from the SDSS DR7 LRG sample 
are robust including systematics of theoretical modeling;
$f(z = 0.3) \sigma_8(z = 0.3) =0.49\pm_{stat.}0.08\pm_{sys.}0.04$, $D_{\rm A} (z = 0.3) =968\pm_{stat.}42\pm_{sys.}17$\,[Mpc], 
$H (z = 0.3) =81.7\pm_{stat.}5.0\pm_{sys.}3.7$\,[km/s/Mpc].
We believe that our method to constrain the cosmological parameters using subhaloes 
catalogues will be useful for more refined samples like CMASS and LOWZ catalogues in the 
Baryon Oscillation Spectroscopic Survey in SDSS-III.

\end{abstract}


\section{INTRODUCTION}
\label{sec:1}
Cosmic acceleration is strongly supported by a recent set of cosmological observations 
including the cosmic microwave background (CMB) anisotropies \citep{Hinshaw:2012lr,Planck-Collaboration:2013fr} 
and Type Ia supernovae \citep{Riess:1998lr,Perlmutter:1999fk,Suzuki:2012jy}. 
Revealing the origin of the cosmic acceleration is one of the key sciences in modern physics,
and there are therefore ongoing or planned cosmological observations from various points 
of view (for a recent review, see \citet{Weinberg:2012uq}). The origin of cosmic acceleration may be 
explained by either of two possible ways as follows. One is to introduce mysterious energy 
component with negative pressure, the so-called dark energy. Another is 
to modify general relativity on infra-red scales while keeping unchanged on small scales 
so that the theory can pass tests of gravity in the solar system. 
It is desirable to establish a methodology to distinguish 
two possibilities and even to identify the nature of dark energy or the theory of gravity on
cosmological scales (for a review of modified gravity, see \citet{Jain:2010fk}).\par 

A standard approach to tackle this problem is to combine measurements of 
the expansion history with those of the growth of the large-scale structure at different time and scales. 
Interestingly, the clustering of galaxies at large scales provides us with a unique 
opportunity to simultaneously measure both probes. 
There are various completed, ongoing or proposed galaxy redshift surveys 
which include the 2dF Galaxy Redshift Survey (2dFGRS) \citep{2dFGRS:Colless:2001qy}, 
Sloan Digital Sky Survey (SDSS) (for the latest data release (DR), see \citet{SDSS-DR10:Ahn:2013kx}), 
WiggleZ dark energy survey \citep{WiggleZ:Blake:2009vn}, 
Subaru Prime Focus Spectrograph (PFS) Survey \citep{Ellis:2012lr}, 
Dark Energy Spectroscopic Instrument (DESI) \citep{DESI:Levi:2013fj}
and Euclid \citep{Euclid:Laureijs:2011jw}. 
Most of such gigantic galaxy redshift surveys are designed to detect
the baryon acoustic oscillation (BAO) scale that is used as a standard ruler 
to explore the expansion history 
\citep{Anderson:2012ap,Beutler:2011kr,Blake:2011kc,Eisenstein:2005lr,Percival:2007cy,Seo:2012la}.\par

While the BAO scale is usually measured from the isotropic or the spherically-averaged 
part of the clustering signal, in this paper we focus on the anisotropic part 
which in fact carries additional information on the cosmic growth and expansion history. 
The anisotropies on the galaxy clustering arise from two effects: 
The first one is the Alcock-Paczynski (AP) effect 
\citep{Alcock:1979kx}. This effect arises if the background expansion 
of the real universe differs from 
the fiducial cosmology when converting the observed galaxy positions, i.e.,
redshift and angular positions, to the comoving radial and transverse 
distances. The measured clustering pattern in distance space
is distorted in a purely geometrical way through this effect. The distortion 
perpendicular to line of sight is proportional to the angular diameter 
distance, $D_{\rm A}(z)$, while one parallel to the line of sight is inversely 
proportional 
to the Hubble parameter, $H(z)$, evaluated with the assumed fiducial cosmology 
\citep{Matsubara:1996zr,Ballinger:1996,Padmanabhan:2008xc}. 
Thus, using the BAO signature imprinted on the galaxy clustering as a standard 
ruler, these distortions can be measured, leading to a precise 
determination of $D_{\rm A}(z)$ and $H(z)$ 
\citep[e.g.][]{Matsubara:2004lr,Seo:2003fk}.
The second effect is the redshift-space distortions (RSDs) caused by peculiar velocities
of galaxies \citep[e.g.][]{Hamilton:1998fk,Peebles:1980qy}. 
Since again the radial positions of galaxies are determined 
by their redshifts, the radial positions in redshift space are contaminated by their peculiar velocities 
along the line-of-sight direction. Thus, RSDs makes the line-of-sight direction special, also inducing 
anisotropies in the apparent clustering. 
In other words, RSDs allow us to extract information on the velocity field through a measurement of
anisotropies, which can be used as a powerful probe of modified gravity because the velocity field is
related to the Newton potential via the Euler equation \citep{Guzzo:2008fr,Yamamoto:2008ys}. 
In the linear theory of density perturbations, RSDs are characterized by the linear growth rate 
defined by $f \equiv d\ln D_{+}(a)/d\ln a$, where $D_{+}(a)$ is the linear growth factor.
Note that this dynamical measure from RSDs is complementary to 
weak lensing analyses or the measurements of the integrated Sachs-Wolfe effect in CMB anisotropies
which probes the sum of two gravitational potentials (e.g. \citet{Bertschinger:2006ys,Kimura:2012aa}).\par 

While RSDs are qualitatively understood as a combination of the squashing effect on large scales, known as 
the Kaiser effect \citep{Sargent:1977fj,Kaiser:1987vn}, and the so-called Finger-of-God (FoG) dilution 
effect at small scale \citep{Jackson:1972lr}, it is challenging to accurately model 
the galaxy clustering at intermediate regime where non-linearity of the structure 
formation cannot be negligible (for a review, see \citet{Bernardeau:2002lr}). 
The difficulty here is the non-linear mapping from real space to redshift space, and 
various efforts have recently been made for the accurate prediction of the power 
spectrum or the correlation function in redshift space on the basis of perturbative approaches 
\citep{Matsubara:2008lr,Matsubara:2008vp,Matsubara:2011qy,
Matsubara:2013lr,Nishimichi:2011xi,Okumura:2012gf,Okumura:2012rq,
Reid:2011kx,Scoccimarro:2004lr,Seljak:2011dw,Taruya:2009hn,Taruya:2010lr,
Taruya:2013qy,Vlah:2012uq,Vlah:2013fj,Wang:2013fk}.
Though such previous works show a successful performance for the clustering of dark matter or 
dark matter haloes when compared with $N$-body simulations, it is still necessary to validate 
such an approach using more realistic galaxy mock catalogues. \par 

The main goal of this paper is to simultaneously obtain robust constraints on $f,D_{A},$ and $H$ from the anisotropic 
galaxy power spectrum. As a specific example, we are going to use the luminous red galaxy (LRG) sample 
in SDSS-II DR7 \citep{SDSSDR7:Abazajian:2009rt,Eisenstein:2001vn}. Even though more recent samples such as
CMASS and LOWZ from Baryon Oscillation Spectroscopic Survey (BOSS) in SDSS-III DR10 
are publicly available \citep{SDSS-DR10:Ahn:2013kx}, it is still interesting to investigate the anisotropic clustering 
signal in the DR7 LRG sample for following reasons. 
First of all, DR7 LRG is one of the samples with which the multipole power spectra are 
properly measured \citep{Hikage:2013zr,Yamamoto:2008ys,Yamamoto:2010kj} with reasonable precisions, 
while some attempts have been made in the literature \citep{Blake:2011fj,Cole:1994aa,Hamilton:1995aa,Hatton:1999aa}. 
In order to measure the redshift-space distortions in the anisotropic distribution of galaxies, 
it is useful to expand the anisotropic power spectrum into the sum of the Legendre polynomials. 
Note also that the significance of the survey window effect in the DR7 LRG catalogue has been discussed 
in \citep{Sato:2013nx,Sato:2011wd}. 

As briefly explained in Section \ref{sec:2}, measuring the multipole power spectra, is not straightforward.
Although one usually assumes a fixed line-of-sight direction in order to utilize the fast-Fourier transformation (FFT), 
\citet{Yoo:2013uq} (and its references therein) shows that this assumption is not appropriate to interpret 
the large-scale anisotropies especially for surveys covering wide sky area such as the SDSS. 
Secondly, a couple of previous works addressed the anisotropic signal of LRGs via the correlation function 
\citep{Samushia:2012ce,Xu:2013fr}, but they did not consider a simultaneous constraint on $f$, $D_{A}$ and $H$
via the AP effect and RSDs. 
As shown in \citet{Taruya:2011kx}, the results could be biased if one of them is artificially fixed 
since there must be moderate degeneracies among $f$, $D_{A}$ and $H$. 
Note that \cite{Reid:2012ly} present such a simultaneous constraint using the DR9 CMASS sample (see also \citet{Chuang:2012uq,Chuang:2013qy,2013MNRAS.435..255C,Hemantha:2013fk} for similar attempts with the DR7 LRGs). \par 

Another reason why there is still room to explore the LRG sample is that no work has tested  
the validity of the methodology to extract the anisotropic signal against a `realistic' mock catalogue. 
In particular, the FoG effect due to satellite galaxies is worrisome even on large scales. 
Studies with high-resolution simulations on dark matter haloes and subhaloes suggest a division of galaxies 
into central and satellite galaxies (e.g. see \citet{Kravtsov:2004fr}). 
The halo occupation distribution (HOD) model (for a review, see \citet{Cooray:2002lr}) is a useful 
tool to describe the link between galaxies and dark matter. Detailed HOD studies via the small-scale 
clustering show that a majority of LRGs is considered as central while roughly $\sim 5$\% of LRGs 
is satellite surrounding central galaxy \citep{Zheng:2008fj,Reid:2009fb}. 
Since satellites have intrahalo velocities causing the FoG effect, it is important to include the effect of 
satellites into the theoretical model when one analyses the anisotropic clustering on smaller scales beyond the limitation of perturbation theory (PT) \citep{Hikage:2013zr}. 
Note that even off-centred central galaxy may also result in additional FoG effect 
\citep{Hikage:2012ss,Hikage:2012uq}. 
Our companion paper shows that a properly-chosen {\it subhalo} catalogue is able to resemble 
such two populations of LRGs and indeed reproduce the measured multipole power spectra 
\citep{Nishimichi:2013lr}. While most of studies with subhaloes are explored in the context of 
subhalo abundance matching, called as the SHAM (e.g. \citet{Conroy:2006fj,Masaki:2012fj}), 
our subhalo catalogue is based on a simple mass cut so that it reproduces the measured 
anisotropic signal and hence complementary to the SHAM scheme.\par 

In this work, we extensively and systematically study our RSD model against such a realistic 
mock catalogue. 
We adopt the PT-based model for non-linear RSDs developed by \citet{Taruya:2010lr}. 
We then combine it with a simple but phenomenological function for linear and scale-dependent 
galaxy bias, motivated by \citet{Nishimichi:2011xi} (see also \cite{Ishikawa:2013yq} for a similar approach). 
After systematically studying the validity of our approach, we will present simultaneous constraints on 
$f,D_{A}$, and $H$ from the DR7 LRG sample. \par 

This paper is organized as follows: Section \ref{sec:2} is devoted to explain the data set we analyse, namely, 
the monopole, quadrupole, and hexadecapole power spectra of the LRG sample in SDSS DR7. 
In Section \ref{sec:3}, we introduce the theoretical model of the multipole power spectra to face with the observation. 
The validity of our analysis is tested against the mock catalogue in Section \ref{sec:4}. 
The results of our analysis with the real data set of the DR7 LRG sample are presented 
and compared with several previous studies in Section \ref{sec:5}. Finally, we summarize this work in Section \ref{sec:6}.

\begin{figure}
\includegraphics[width=80mm]{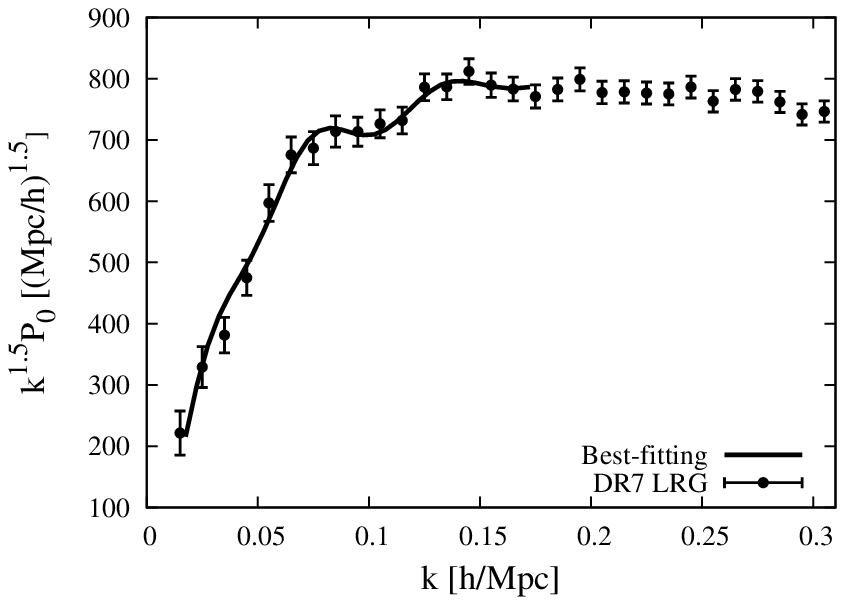}
\includegraphics[width=80mm]{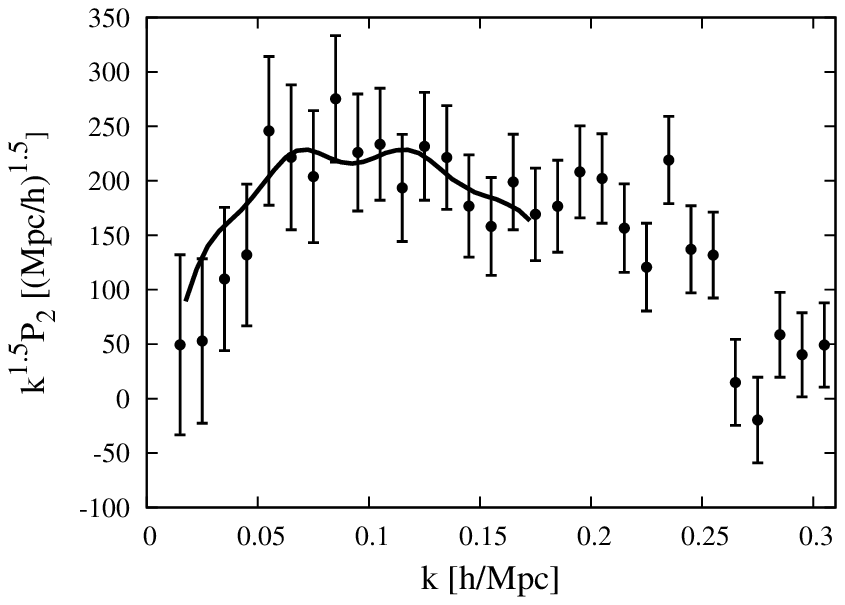}
\includegraphics[width=80mm]{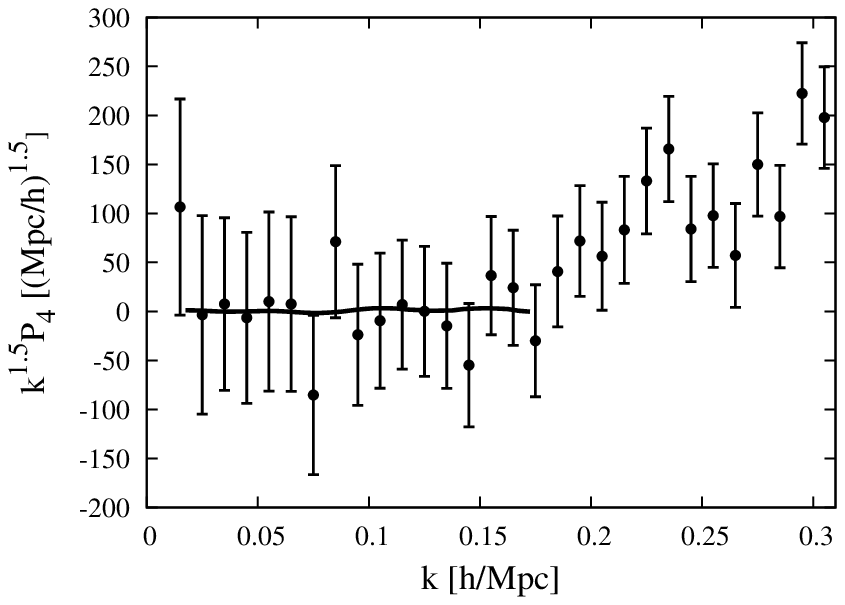}
\caption{The filled circles
with error bars are the observed multipole spectra, 
monopole (top), quadrupole (middle), and hexadecapole 
(bottom) power spectra of the SDSS DR7 LRG sample.  
We plot the best-fitting results with solid curves, whose details are 
described in Section \ref{sec:5}. The results are 
multiplied by $k^{1.5}$. 
The best-fitting curves are plotted in the range of the wavenumbers 
$k\leq k_{\rm max} = 0.175 [h$/Mpc] that corresponds to the valid range 
of our theoretical model (see Section \ref{sec:4.3}). 
We used the data 
in the range of the wavenumbers $k\leq k_{\rm max} = 0.175[h$/Mpc], 
which include 51 data points, as described in Section \ref{sec:5}.}
\label{fig:fitSDSSpower}
\end{figure}

\section{POWER SPECTRUM MEASUREMENT}
\label{sec:2}
In this section, we briefly summarize the measurement of the multipole 
power spectra of LRGs in the SDSS DR7 \citep{SDSSDR7:Abazajian:2009rt}.
We use the same data set as in \citep{Yamamoto:2010kj}. 
The DR7 LRG sample is selected to cover the redshift range, 
$0.16 < z < 0.47$, only in the northern cap in order to reduce systematics 
uncertainties. Thus, the sky coverage is limited to 
$\sim 7,150$ deg$^2$ and the total number of LRGs is $N_{\rm LRG}=100,157$. 
This corresponds to a survey volume of $V_{\rm survey}\sim1.3$ [Gpc$^3/h^3$].
We adopt the same method for the measurement as in \citep{Yamamoto:2010kj,Yamamoto:2006sp} but with 
the fiducial cosmological background favored by the Planck results \citep{Planck-Collaboration:2013fr}.
Namely, we adopt the distant-redshift 
relation of the spatially flat $\Lambda$CDM cosmology with 
$\Omega_{\rm m}=0.32$, $h=0.67$.

In our algorithm, the line-of-sight direction is chosen 
for each pair of galaxies, which enables us to measure 
the higher multipole spectra without introducing the 
fixed line-of-sight direction.
With this algorithm, the window effect can be kept small for the SDSS DR7 
LRG sample, which we neglect in the present paper \citep{Sato:2011wd}. 

The multipole power spectra, $P_\ell(k)$, are the coefficients of 
the Legendre multipole expansion of the anisotropic power spectrum,
\begin{equation}
P^{s}(k,\mu)=\sum_{\ell={\rm even}}P_{\ell}(k)\L_{\ell}(\mu),
\end{equation}
\begin{equation}
P_{\ell}(k) = \frac{2\ell + 1}{2}\int_{-1}^{1}d\mu \ \L_{\ell}(\mu)P^{s}(k, \mu),
\end{equation}
where $\L_{\ell}(\mu)$ is the $\ell$-th Legendre polynomial, and $\mu$ is 
the directional cosine between the line-of-sight direction and the wavenumber vector. 
The anisotropic power spectrum is useful to constrain both the 
cosmic expansion history and the growth history of the large 
scale structure of the universe. In the present paper, we 
demonstrate the cosmological constraints from the combination of the
monopole, quadrupole, and hexadecapole power spectra, $P_0(k)$, $P_2(k)$, 
and $P_4(k)$, which provide almost as much information on $f, D_{\rm A}$, and $H$ as 
the full two-dimensional anisotropic power spectrum as long as one restricts the
analysis within the valid range of perturbative approaches \citep{Taruya:2011kx}.

In our measurement of the multipole power spectra, we adopt an
estimator, $\hat{P}_{\ell}(k)$, for discrete density fields \citep{Yamamoto:2006sp}:
\begin{equation}
\hat{P}_{\ell}(k)=\frac{2\ell+1}{\Delta V_k}\int_{\Delta V_k}d^3k \ 
\{R_{\ell}(\k)-S_{\ell}(\k)\},
\end{equation}
where $\Delta V_k$ is the volume of a shell in the Fourier space, 
$R_{\ell}$ and $S_{\ell}$, are defined by
\begin{align}
R_{\ell}(\k)=A^{-1}&\Big[\sum_{i_1}^{N_{\rm LRG}}e^{i\k\cdot\s_{i_1}}\L_{\ell}(\mu_{i_1})
-\gamma\sum_{j_1}^{N_{\rm rand}}e^{i\k\cdot\s_{j_1}}\L_{\ell}(\mu_{j_1})\Big] \notag \\
& \times \Big[\sum_{i_2}^{N_{\rm LRG}}e^{i\k\cdot\s_{i_2}}
-\gamma\sum_{j_2}^{N_{\rm rand}}e^{i\k\cdot\s_{j_2}}\Big],
\end{align}
\begin{equation}
S_{\ell}(\k)=A^{-1}(1+\gamma)\sum_{i}^{N_{\rm LRG}}\L_{\ell}(\mu_{i}),
\end{equation}
where $\s_{i}$ and $\s_{j}$ are positions of galaxies 
and of random samples in redshift space, respectively, 
$\gamma=N_{\rm LRG}/N_{\rm rand}$, is the ratio of 
the number of LRGs to that of random samples, for which we set 
$\gamma=0.05$, and  $\mu=\k\cdot\s/|\k||\s|$. 
Here $A$ is defined by the integration of the mean number density over 
observed redshift,
\begin{equation}
A=\Delta\Omega\int^{s(z_{\rm max})}_{s(z_{\rm min})}dss^2 \ \bar{n}^2(z),
\end{equation}
where $s$ is the radial (comoving) coordinate of the fiducial cosmological background, 
and $\Delta \Omega$ is the solid angle of the survey area.

The statistical error of the multipole power spectra may be estimated by the formula in \citep{Yamamoto:2006sp}.
Adopting the constant weight factor, we have
\begin{align}\label{eq:error}
&\Delta \hat{P}_{\ell}(k)^2 = 2(2\ell+1)^2\frac{(2\pi)^3}
{\Delta V_{k}^{2}}\int_{\Delta V_k} d^3k \ \frac{1}{A^2} \notag \\
& \ \times\int^{s(z_{\rm max})}_{s(z_{\rm min})}ds \ 
\bar{n}^4(z)[P(\k,\s)+\frac{1}{\bar{n}(\s)}]^2
\L_{\ell}^{2}(\hat{\s}\cdot\hat{\k}),
\end{align}
where we used the approximation $P(\k,\s)\simeq P_0(k)+P_2(k){\cal L}_2(\hat{\s}\cdot\hat{\k})$, 
which has been derived on the basis of the so-called FKP method \citep{Feldman:1994ys}. 

This estimation of the statistical error is not strict but rather optimistic in the following points.  
First, covariances between different $\ell$-th multipoles are neglected.
Second, covariances between different $k$-bins, i.e., the non-Gaussian error, 
from the window effect and the non-linear gravitational growth are not taken into account. 
Our cosmological parameter estimation in the following sections 
is slightly altered when the covariances between different $\ell$- and $k$-bins are properly taken into account,
which are obtained in \citep{Sato:2013nx,Sato:2011wd}. 
However, the effect of such non-Gaussian error on the resultant one-dimensional 
marginalized errors is small, as shown in \citet{Takahashi:2009kl,Takahashi:2011lr}. 
Therefore we will consider only the diagonal component of the covariance matrix for simplicity from now on. 

Figure~\ref{fig:fitSDSSpower} demonstrates the resultant multipole power spectra,
the monopole (top), quadrupole (middle), and hexadecapole (bottom), respectively. 
The solid curves in each panel show the best-fitting results described in Section \ref{sec:5}.

\section{MODELING THE MULTIPOLE POWER SPECTRA}
\label{sec:3}
In this section, we briefly review the theoretical model of the multipole power spectra used 
in the cosmological analysis. Our goal is to constrain the linear growth rate 
and geometrical factors 
simultaneously through RSDs and AP effect in an unbiased manner. For this purpose, a proper 
modeling of the shape and the amplitude of the anisotropic power spectrum is 
rather crucial (e.g. \citet{Padmanabhan:2008xc}), 
and we will investigate the robustness of our model in detail in Section \ref{sec:4}. 
The model presented here is based on the perturbation theory calculation, 
and we will separately give prescription on how to 
compute the multipole power spectra.

\subsection{Redshift-space distortions and Non-linear gravitational growth}
\label{sec:3.1}

Redshift-space distortions and gravitational clustering involve, in nature,  
non-linear and non-Gaussian effects, and it is quite essential to take 
a proper account of these for a robust cosmological analysis beyond the linear scales. 
Since we are 
interested in a large-scale anisotropic clustering at moderately high redshift, 
the PT approach should work well, and 
a percent-level precision
is achievable with PT calculation in weakly non-linear regime 
$k\lesssim0.2\,[h/$Mpc]. 

Let us first consider RSDs. It is well-known that 
the clustering statistics in redshift space are influenced by the 
two effects, the Kaiser and Finger-of-God effects. While the former comes from 
the coherent motion of galaxies and enhances the clustering 
amplitude, the latter is mainly attributed to the virialized random motion 
of galaxies sitting in a halo and suppresses the power spectrum significantly 
along the line of sight. Strictly speaking, these effects cannot be
treated separately, and through the higher-order corrections, 
a tight correlation between the density and velocity fields still 
plays an important role on the scales of our interest. In the present paper, 
among several proposed models to account for 
the non-linear RSDs (\citep{Matsubara:2008vp,Reid:2011kx,Seljak:2011dw}),
we adopt the model given by \citep{Taruya:2010lr} (hereafter, TNS model):  
\begin{align}\label{eq:TNS}
P^{s}(k,\mu) &= D_{{\rm FoG}}(k\mu f\sigma_{\rm v}) \notag \\
\, &\times \Big[P_{{\rm Kaiser}}(k,\mu;\,f) + 
A(k,\mu;\,f) + B(k,\mu;\,f) \Big],
\end{align}
where $\sigma_{\rm v}$ is a nuisance parameter, which is related 
to the one-dimensional velocity dispersion. The function 
$D_{{\rm FoG}}(k\mu f\sigma_{\rm v})$ 
characterizes the suppression of the power spectrum by the FoG effect, 
for which we adopt the Gaussian form;
\begin{equation}\label{eq:FoG}
D_{{\rm FoG}}(x) = \exp(-x^2).
\end{equation}
The function, $P_{{\rm Kaiser}}(k,\mu)$, is the non-linear generalization of the 
Kaiser term given by \citep{Scoccimarro:2004lr}
\begin{equation}\label{eq:Kaiser}
P_{{\rm Kaiser}}(k,\mu;\,f) = P_{\delta \delta}(k) + 2f\mu^2P_{\delta \theta}(k) + f^2\mu^4P_{\theta \theta}(k).
\end{equation}
Here, the functions $P_{\delta \delta}(k)$, $P_{\theta \theta}(k)$, and 
$P_{\delta \theta}(k)$ are respectively the auto-power spectra 
of the density and the velocity divergence, and their cross-power spectrum. 
Here, the velocity divergence, $\theta$, is normalized as $\theta\equiv-\nabla \v/(faH)$.

The main characteristic of the model (\ref{eq:TNS}) is the two additional 
terms $A$ and $B$, which represent the higher-order coupling between 
the velocity and density fields, usually ignored in a phenomenological 
model of RSDs. These corrections have been properly derived 
on the basis of the low-$k$ expansion from the exact expression of the anisotropic power spectrum, expressed as
\begin{align}
A(k,\mu;\,f)&=(k\mu\,f)\,\int \frac{d^3\p}{(2\pi)^3} \,\,\frac{p_z}{p^2} 
\nonumber \\
&\times \Big[B_\sigma(\p,\k-\p,-\k)-B_\sigma(\p,\k,-\k-\p)\Big],
\label{eq:Aterm}
\\
B(k,\mu;\,f)&= (k\mu\,f)^2\int \frac{d^3\p}{(2\pi)^3} F(\p)F(\k-\p);
\nonumber
\\
&F(\p)=\frac{p_z}{p^2}
\Big[ P_{\delta \theta}+f\,\frac{p_z^2}{p^2}\,P_{\theta \theta}\,\Big],
\label{eq:Bterm}
\end{align}
where the function, $B_\sigma(\k_1,\k_2,\k_3)$, is the 
cross bispectrum defined by
\begin{align}
&(2\pi)^3\dD(\k_1+\k_2+\k_3)\,B_\sigma(\k_1,\k_2,\k_3)
\nonumber\\
&\quad
=\Big\< \theta(\k_1)
\Big[\delta(\k_2)+f\,\frac{k_{2z}^2}{k_2^2}\theta(\k_2)\Big]
\Big[\delta(\k_3)+f\,\frac{k_{3z}^2}{k_3^2}\theta(\k_3)\Big]
\Big\>.
\label{eq:def_B_sigma}
\end{align}
It is shown in the previous study that these two terms enhance 
the amplitude of the power spectrum over the wavenumbers 
where the baryon acoustic feature is prominent, 
and moderately but notably change the 
acoustic structure imprinted on the power spectrum \citep{Taruya:2010lr}. 
As a result, the model (\ref{eq:TNS}) successfully describes both the  
matter and halo power spectra of $N$-body simulations 
at weakly non-linear scales. 
In particular, the non-Gaussian contribution described by $A$ term exhibits a strong 
dependence on halo/galaxy biasing, and 
in addition to the linear Kaiser effect, it gives a rather prominent
enhancement in the multipole power spectra \citep{Nishimichi:2011xi}. 
Since the effect is known to be significant for highly biased objects, 
the model (\ref{eq:TNS}) seems best suited for characterizing 
the anisotropic LRG clustering in weakly non-linear regime.

To compute the power spectrum (\ref{eq:TNS}), we need to further 
incorporate the effect of non-linear gravitational growth into each term. 
In this paper, we apply the resummed PT scheme called 
{\tt RegPT} \citep{Taruya:2012ai}, and following the prescription described 
in \citep{Taruya:2013qy}, we evaluate the power spectrum and 
bispectrum contributions, consistently including the non-linear corrections 
up to the two-loop order, i.e., next-to-next-leading order.  
The {\tt RegPT} scheme is based on the multipoint propagator expansion, 
in which non-perturbative properties of gravitational growth  are wholly 
encapsulated. With this scheme, 
any statistical quantities consisting of the density and velocity fields are 
built up with the multipoint propagators.
Making use of the analytic properties of 
the propagators, a novel regularized treatment has been implemented, 
which allows us to consistently reproduce the standard PT results at 
low-$k$ and the expected resummed behavior at high-$k$. 
It has been demonstrated that
the proposed scheme can be used to give a percent-level prediction 
of the power spectrum and the correlation function at weakly non-linear regime 
in both real and redshift spaces \citep{Taruya:2012ai,Taruya:2013qy}. At redshift $z\simeq0.3$, while 
the standard PT fails to reproduce the matter power spectrum at 
$k\sim0.1\,[h$/Mpc], the applicable range of {\tt RegPT} is rather 
wider, and it can cover the almost entire scales of BAOs.

\begin{figure*}
\includegraphics[width=80mm]{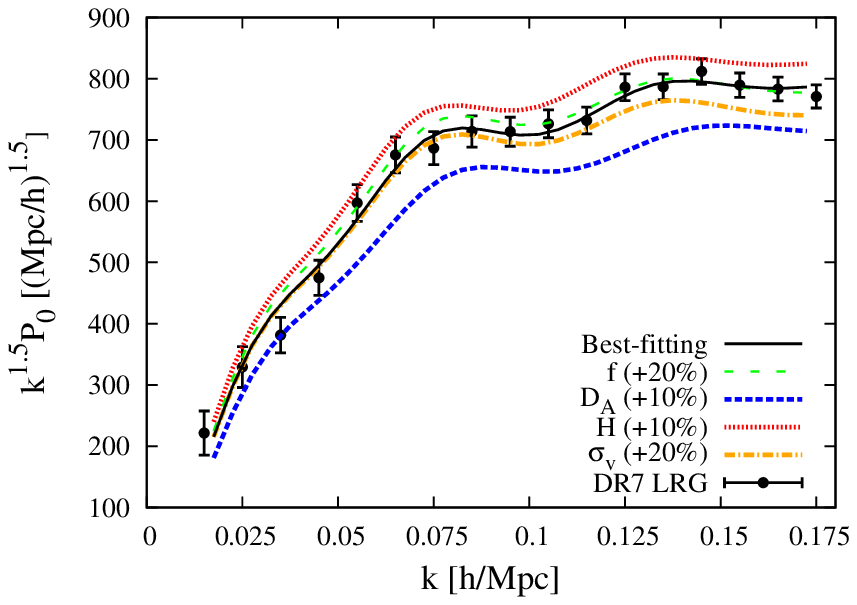}
\includegraphics[width=80mm]{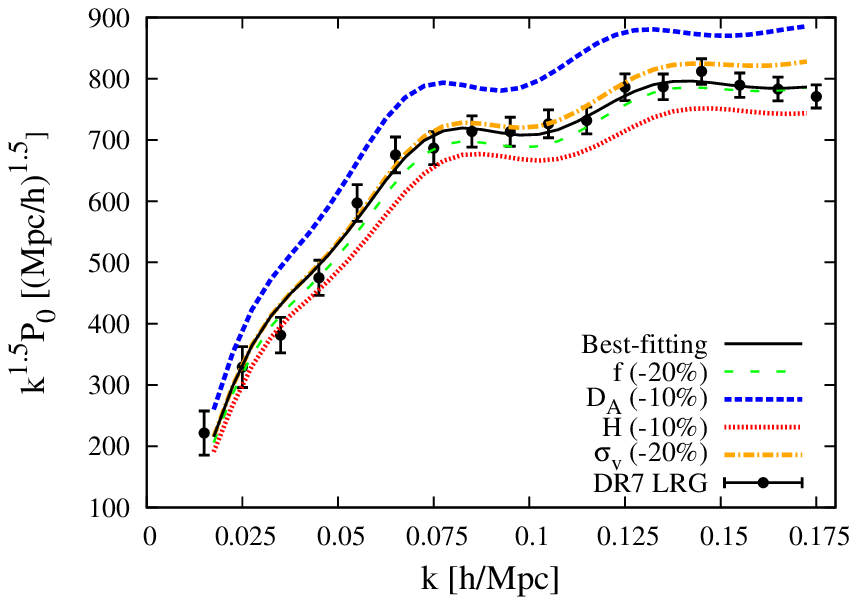}

\includegraphics[width=80mm]{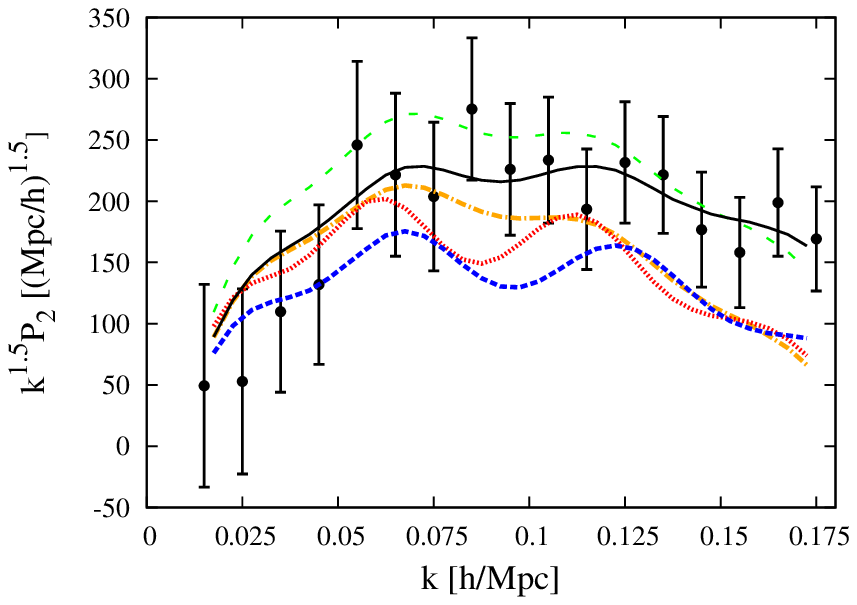}
\includegraphics[width=80mm]{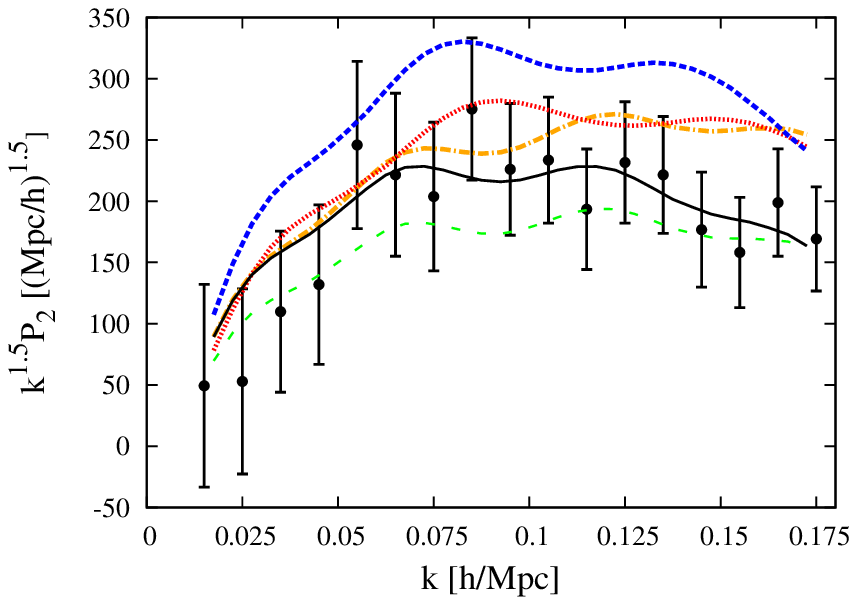}

\includegraphics[width=80mm]{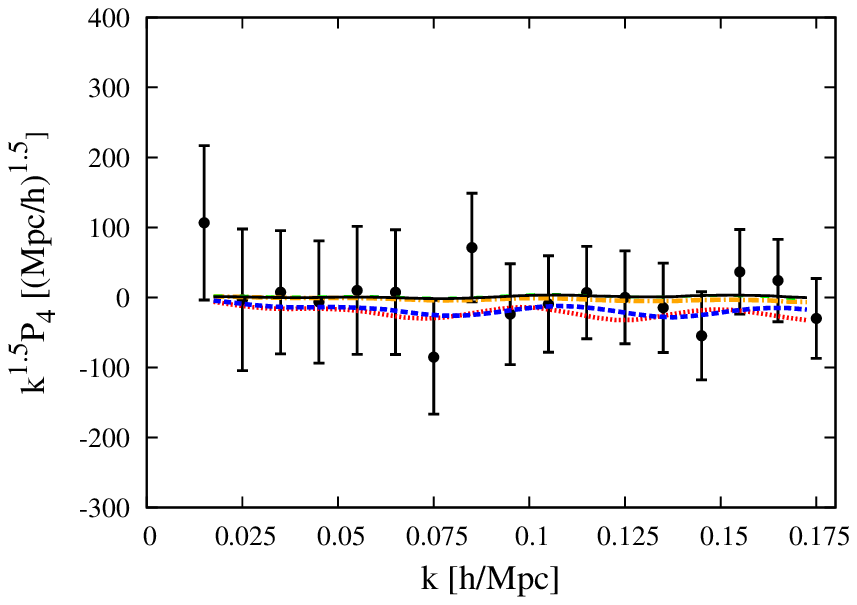}
\includegraphics[width=80mm]{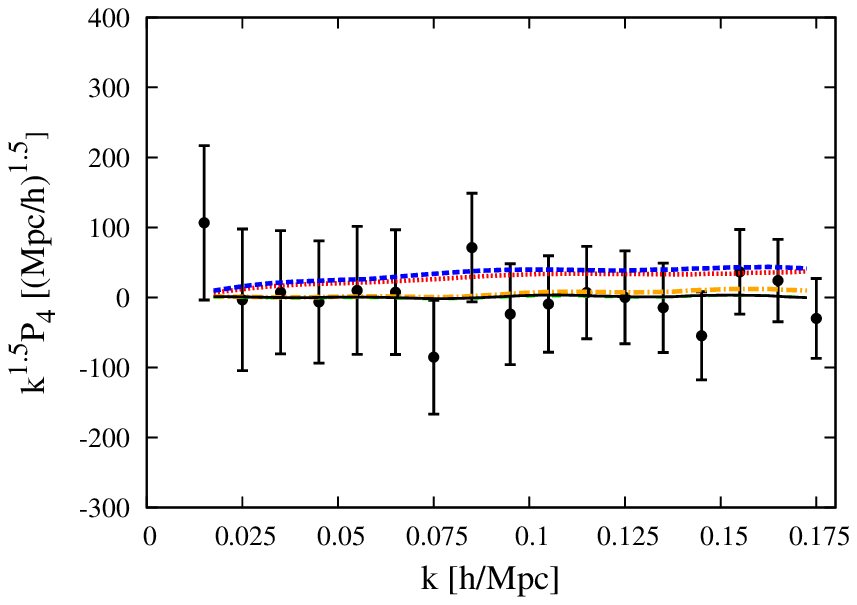}
\caption{
Variations of monopole (top), quadrupole (middle), and hexadecapole 
(bottom) power spectra computed with the PT model.
We plot the best-fitting model of the SDSS DR7 LRG 
sample in solid curves (see Section~\ref{sec:5} and Table.~\ref{tab:systematic}), 
but the other curves are the models with slightly shifting (left: increasing, 
right: decreasing) parameters, which characterize the linear Kaiser, FoG, and AP effects: 
$f$ ($\pm20\%$: (green; colors are available for the online version) dashed), 
$D_{\rm A}$ ($\pm10$\%: (blue) dotted), 
$H$ ($\pm10\%$: (red) short-dotted), 
and $\sigma_{\rm v}$ ($\pm20\%$: (orange) dot-dashed). 
For reference, the measured power spectra are also plotted by
filled circles with error bars.
}
\label{fig:modelpower}
\end{figure*}

\subsection{Galaxy bias}
\label{sec:3.2}

With the {\tt RegPT} scheme, the model (\ref{eq:TNS}) can give us an accurate 
prediction for the multipole power spectra with its applicable range 
much beyond linear scales. Note, however, that 
the aforementioned model has been originally proposed for 
the matter distribution, and a proper account of the galaxy bias is 
necessary for a cosmological analysis with the DR7 LRG sample.
While there have been 
several sophisticated PT schemes proposed to simultaneously 
characterize the galaxy bias, 
redshift-space distortions, and non-linear gravitational growth, we 
here adopt a rather simple approach. Namely, we assume a 
linear bias relation, similar to the previous study with halo clustering 
\citep{Nishimichi:2011xi}, allowing to incorporate the scale dependence of bias 
into the model, equation~(\ref{eq:TNS}), through the following relation: 
\begin{equation}\label{eq:bias}
\delta_{\rm g}(k)=b(k)\delta_{\rm m}(k).
\end{equation}
With this relation, the expression (\ref{eq:TNS}) is 
replaced with 
\begin{align}
\label{eq:TNS_linear_bias}
P^{s}_{\rm g}(k,\mu) &= D_{{\rm FoG}}(k\mu f\sigma_{\rm v})\notag \\
& \, \, \times\,b(k)^2\,
\Big[P_{{\rm Kaiser}}(k,\mu;\,\beta) + 
b\,A(k,\mu;\,\beta) + b^2\,B(k,\mu;\,\beta) \Big],
\end{align}
where the quantity $\beta$ is defined by $\beta=f/b$. 
For the function $b(k)$, we adopt the following parameterized form: 
\begin{equation}\label{eq:powerlaw}
b(k)=b_0\frac{1+A_2k^2}{1+A_1k}. 
\end{equation}
Note that the functional form (\ref{eq:powerlaw}) is quite close to the one
introduced in \citet{Cole:2005uq}.

The scale-dependent linear bias 
has also been used to describe the redshift-space halo clustering.
Adopting the parameterized function simular to equation~(\ref{eq:powerlaw}), 
the model (\ref{eq:TNS}) successfully describes the multipole power spectra 
of haloes \citep{Nishimichi:2011xi}. In this respect, albeit a rather 
phenomenological treatment, the model of equation~(\ref{eq:TNS}) with 
equation~(\ref{eq:bias}) provides a practically useful prescription, and it is 
worth further testing the robustness with the mock subhalo catalogue described in the next 
section.

\subsection{Alcock-Paczynski effect} 
\label{sec:3.3}

Finally, the remaining effect to be incorporated into the 
theoretical template is the Alcock-Paczynski effect. The AP effect 
arises from the apparent mismatch of the underlying cosmology when 
we convert the redshift and angular position of each galaxy 
to the comoving radial and transverse distances, and it  
modulates the shape and amplitude of the multipole power spectra. 
Using the BAO scale as a standard ruler, this geometrical effect offers an 
attractive method to measure the angular diameter distance $D_{\rm A}(z)$ 
and Hubble parameter $H(z)$ of distance galaxies at redshift $z$. 

The anisotropies caused by the AP effect can be modeled into 
the anisotropic power spectrum as
\begin{equation}\label{eq:AP}
P^{{\rm model}}(k,\mu)=\frac{H}{H^{{\rm fid}}}\Big(\frac{D_{A}^{{\rm fid}}}{D_{\rm A}}\Big)^2P^{s}_{\rm g}(q,\nu).
\end{equation}
Here, the comoving wavenumber $k$ and the directional cosine $\mu$ 
measured with the underlying cosmological model are related to 
the true ones, $q$ and $\nu$, defined by 
\begin{equation}\label{eq:q}
q(k,\mu)\equiv\alpha(\mu) k,
\end{equation}
\begin{equation}\label{eq:nu}
\nu(k,\mu)\equiv\frac{1}{\alpha(\mu)}\frac{H}{H^{{\rm fid}}}\mu,
\end{equation}
where the function $\alpha(\mu)$ is defined as
\begin{equation}\label{alpha}
\alpha(\mu)\equiv\sqrt{\mathstrut \Big(\frac{D_{A}^{{\rm fid}}}{D_{A}}\Big)^2 
+ \Big[\Big(\frac{H}{H^{{\rm fid}}}\Big)^2-\Big(\frac{D_{A}^{{\rm fid}}}{D_{A}}\Big)^2\Big]\mu^2}.
\end{equation}
The quantities $D_{\rm A}^{\rm fid}$ and $H^{\rm fid}$ are
the fiducial values of the angular diameter distance and the Hubble 
parameter at a given redshift slice.
Summing up all the ingredients, we model the multipole power spectra of the DR7 LRG as
\begin{equation}
 P_{\ell}^{\rm model}(k) = \frac{2\ell+1}{2}\int^{1}_{-1}d\mu\,\L_{\ell}(\mu)\,P^{\rm model}(k,\mu). 
 \label{eq: P_ell^model}
\end{equation}

In Fig.~\ref{fig:modelpower}, we plot 
the multipole power spectra computed with our model, and show how the 
ingredients incorporated into our model change the shape and amplitude 
of the monopole (top), quadrupole (middle), and hexadecapole spectra (bottom).  
Here the solid curve is the best-fitting model of the DR7 LRG sample obtained in Section \ref{sec:5} 
(see Table.\ref{tab:systematic}), and the other curves are the models shifting the each parameter
$f$, $D_{\rm A}$, $H$, and $\sigma_{\rm v}$ by $\pm10$-$20$\% (see figure legend), 
while fixing others as well as the bias parameters 
($b_0=2.03$, $A_1=-0.615$, and $A_2=0.0392$). A slight increase 
in the linear growth rate $f$ basically alters the quadrupole-to-monopole ratio 
through the linear Kaiser effect, and the FoG effect characterized 
by $\sigma_{\rm v}$ leads to a suppression of the amplitude of the spectra, 
especially in the quadrupole at the scale of our interest. On the other hand, 
the geometric quantities $D_{\rm A}$ and $H$ 
change not only the acoustic scales in the monopole but also the shape and 
amplitude of the higher multipole spectra through the AP effect. Notably, the AP 
effect changes the quadrupole spectrum significantly. These characteristic 
behaviors indeed play an important role in the 
cosmological parameter estimation, and are
the keys to robustly get simultaneous constraints on $f$, $D_{\rm A}$, and $H$.

\begin{figure}
\includegraphics[width=80mm]{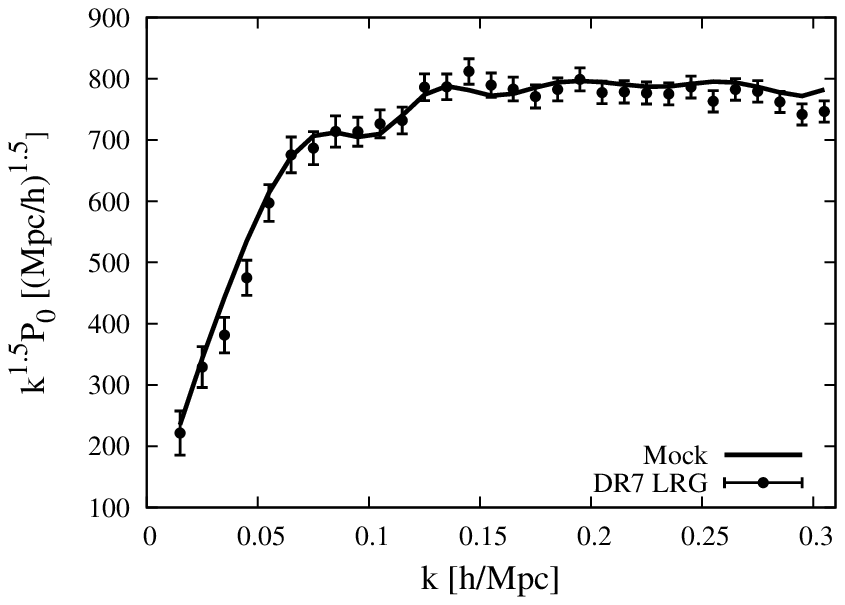}
\includegraphics[width=80mm]{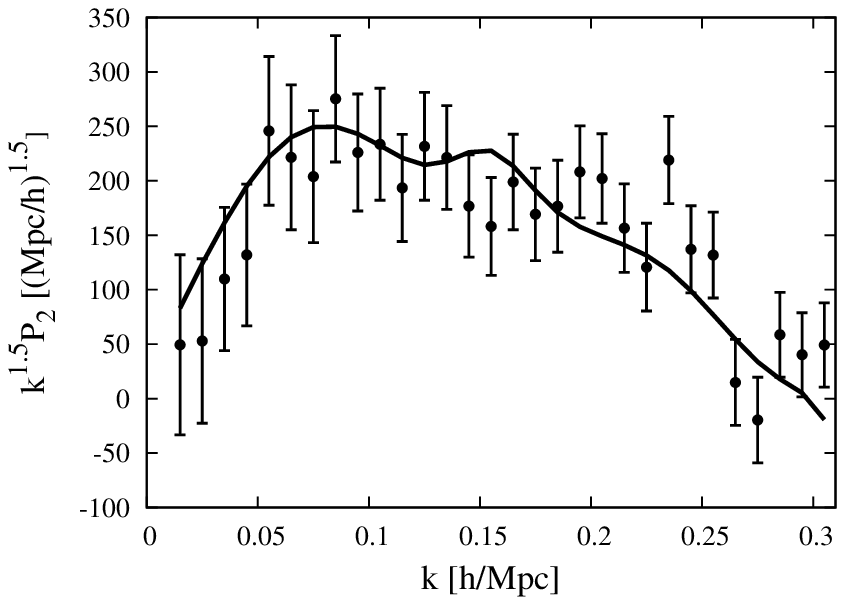}
\includegraphics[width=80mm]{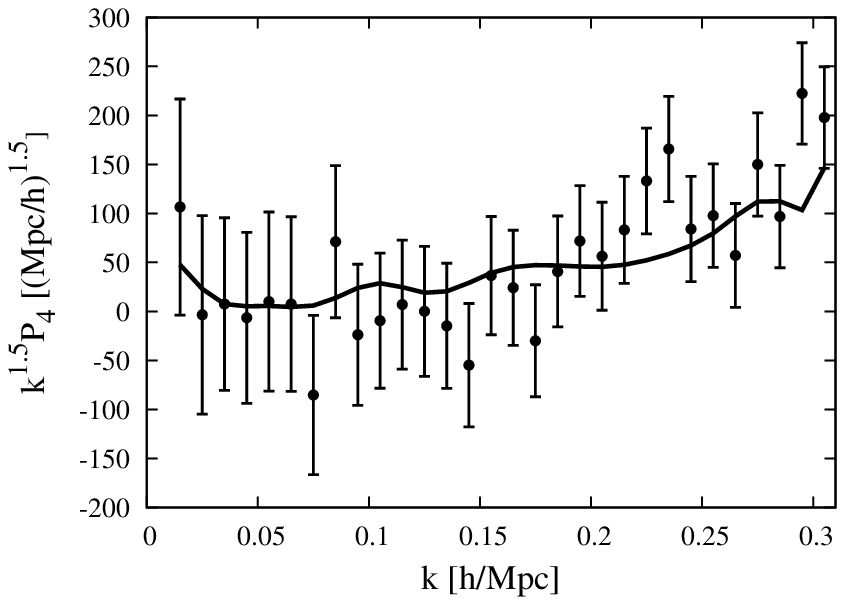}
\caption{
Monopole (top), quadrupole (middle), and hexadecapole (bottom) power spectra measured from our best-fitting 
mock catalogue and those of the SDSS DR7 LRG sample. The black solid lines 
correspond to our mock subhalo catalogue. The filled circles with error bars correspond to the SDSS DR7 LRG sample.
}
\label{fig:mockpower}
\end{figure}

\section{TESTING PT MODELS AGAINST MOCK CATALOGUES}
\label{sec:4}

In this section, we extensively examine the robustness of our model described in the previous section, 
and investigate in detail the applicable range of wavenumbers and the limitation of the model in order to 
correctly estimate the parameters $f$, $D_{\rm A}$, and $H$. 
A great emphasis here is that we test the PT model against `realistic' 
mock catalogues constructed with subhaloes identified from $N$-body simulations. 
There are several popular methods in the literature to construct mock 
catalogues, and these have been applied to characterize 
the observed properties of LRGs. 
The methods include the HOD modeling  
\citep{Brown:2008qy,Kulkarni:2007ve,Reid:2009rr,Reid:2009fb,White:2007dp}, 
and the subhalo abundance matching (SHAM) scheme \citep{Conroy:2006fj,Masaki:2012fj}. 
These mock subhalo catalogues are constructed such that they reproduce the observed number density, 
multiplicity function that determines the number distribution of LRGs hosted by the same halo, 
or the (angular) clustering on relatively small scales.
In contrast, we construct mock catalogues on the basis of the anisotropic clustering pattern of LRGs.
To be more specific, we vary the threshold mass of haloes above which they can host LRGs ($M_{\rm min}^{\rm host}$), 
the minimum mass of subhaloes for LRG candidates ($M_{\rm min}^{\rm sub}$), 
and the fraction of satellite LRGs ($R_{\rm S}$) and search for
the parameters that best reproduce the observed multipole power spectra 
using the Markov-Chain Monte-Carlo (MCMC) method. 
This method is suitable for our purpose 
because the anisotropic clustering pattern 
is in general sensitive to the location and motion of galaxies, and these 
are thought to be tightly related to their size or mass 
along with the merger history of (sub)haloes.

In particular, subhaloes can exhibit a strong FoG 
effect in the measured multipole spectra due to the virial motion 
of satellites, i.e., less massive subhaloes away from the centre of each halo.  
We briefly explain how to generate our mock catalogues and refer to 
the accompanying paper \citep{Nishimichi:2013lr} for more details on the mock construction.

In the latter part of this section, using these catalogues whose input cosmological parameters are 
a priori known by definition, we systematically examine our procedure to estimate the cosmological 
parameters of interest. Our mock test shows a successful performance, and hence validates our 
theoretical template and methodology which will be applied to the real data in the following section.

\subsection{Generation of mock catalogues}
\label{sec:4.1}

We begin by describing the cosmological $N$-body simulations used in our analyses. 
We use $11$ independent random realizations of cosmological $N$-body simulations
presented in \citep{Nishimichi:2011xi}, which are created by a publicly-available $N$-body code, 
{\tt Gadget2} \citep{Springel:2005qe}.
Each realization includes $N = 1,280^3$ dark matter particles in a cubic box with a side length of 
$1,144.72$ [Mpc/$h$], which result in a particle mass of $5.54\times10^{10}$[$M_\odot/h$].
They assume a flat $\Lambda$CDM model with the cosmological parameters 
consistent with the five-year WMAP results \citep{Komatsu:2009ly}; 
$\Omega_{\rm m} = 0.279$, $\Omega_{\rm b} = 0.046$, $h=0.701$, $n_{\rm s} = 0.96$ and $\sigma_8=0.817$. 
The initial conditions are generated at $z_{\rm in} = 99$ using 
an initial condition generator developed 
in \citep{Nishimichi:2009lq,Valageas:2011tg} 
based on the second-order Lagrangian perturbation theory 
\citep{Crocce:2006kx,Scoccimarro:1998eu}. 
Haloes and subhaloes are respectively identified with 
{\tt Friends-of-Friends} (FoF; e.g. \citet{Davis:1985vn}) 
and {\tt SUBFIND} \citep{Springel:2001oq} algorithm 
from the dark matter positions and velocities at $z_{{\rm out}}=0.35$. 
In each halo, we distinguish the most massive subhalo from the rest, 
and conventionally call the most massive one {\it central}, 
while other remaining subhaloes are called {\it satellites}. 
Collecting the subhaloes that satisfy 
the criteria set by the three parameters (i.e., minimum mass of host haloes $M_{\rm min}^{\rm host}$, 
minimum mass of subhaloes $M_{\rm min}^{\rm sub}$ 
and the fraction of satellite subhaloes $R_{\rm S}$), 
we record their centre-of-mass positions and velocities to form mock LRGs
(this catalogue is called as Model 4a in \citet{Nishimichi:2013lr}).

Provided the mock LRGs, the multipole moments of their power spectrum are measured as follows.
We first evaluate the density field of subhaloes assigned on $1,024^3$ regular grid points by Nearest-Grid-Point 
interpolation technique (NGP). 
We then transform it into the Fourier space, and correct the window function by dividing by the NGP window kernel. 
Finally, with an appropriate weight depending on the directional cosine
of the wavevector (i.e., the Legendre polynomial, $\mathcal{L}_\ell(\mu)$),  
the density squared is fitted with the cubic B-spline function as function of wavenumber, 
from which the multipole spectra are 
evaluated at the wavenumbers where the observed power spectra are given. 
This method has an advantage over the standard power spectrum 
estimation with binned wavenumbers 
in the sense that the effect of finite number of grids over $\mu$,  
which significantly affects the estimation of higher multipole spectra, 
is greatly reduced with cubic B-spline function. 
Note that the effect of finite grid size cannot be simply mitigated by 
increasing the number of simulations unless one changes the box size. 

These above steps (i.e., construction of subhalo catalogues and 
estimation of the power spectrum) are repeated with different set of parameters,  
($M_{\rm min}^{\rm host}$, $M_{\rm min}^{\rm sub}$, $R_{\rm S}$). 
We compare the resultant power spectra with the observation to 
find the best-fitting mock catalogue to the DR7 LRGs. We carried out this 
with the MCMC algorithm. The properties of the best-fitting catalogue are 
discussed in the next subsection.

\subsection{Properties of a best-fitting mock LRG catalogue}
\label{sec:4.2}
We use the best-fitting catalogue obtained in the accompanying paper \citep{Nishimichi:2013lr}. 
The best-fitting parameters for this catalogue are $M_{\rm min}^{\rm host} = 9.81\times10^{12}M_{\odot}/h$, 
$M_{\rm min}^{\rm sub} = 8.86\times 10^{12}M_{\odot}/h$, $R_{\rm S}=0.26$ with $\chi^{2}$/d.o.f. = 0.87, 
where d.o.f. = 30$\times$3 - 3 = 87 is the total number of data points of $P_0(k_i)$, $P_2(k_i)$ and $P_4(k_i)$ 
in the range of the wavenumbers $k_i<k_{\rm max} = 0.305\, [h/$Mpc] (30 points for each) minus the number of 
free parameters (3). 
The multipole power spectra of this catalogue are shown in solid lines in Fig.~\ref{fig:mockpower}. 
Overall, the power spectra measured from our mock catalogue reproduce the three observed moments, and 
the value of $\chi^2$ indeed suggests that our mock is statistically consistent with the observation.

The satellite fraction of 26 \%, which is significantly larger than that suggested from the 
Count-in-Cylinder (CiC) analysis (5 \%; \citet{Reid:2009fb}), mainly comes from the strong damping of 
the observed quadrupole moment towards high-$k$ while successfully reproducing the amplitude of the 
monopole moment at the same time. 
These two results seem contradictory each other, but they would be understood as follows.
The higher satellite fraction 
we found originates from the kinematical feature in RSDs, while the CiC analysis identified close galaxy 
pairs such that they both sit in the same massive halo. In other words, roughly 20\% of satellite 
in our terminology could just correspond to a single LRG system in which the central galaxy is not 
observed or it has a significant off-centring (see \citet{Nishimichi:2013lr} in detail.)

The importance of the satellite fraction is also reported by \citet{Hikage:2013zr}, in which the authors
discuss its effect on the higher multipole spectra of LRGs in redshift space.
Since satellite galaxies have a velocity structure different from that of centrals, 
one can alter the multipole spectra, especially higher multipoles, 
by changing their fraction even when the HOD is kept unchanged.
A larger satellite fraction means a larger velocity dispersion, and it leads to a stronger FoG damping
required to explain the observed quadrupole.

\subsection{Test of systematics with the mock catalogue}
\label{sec:4.3}

In what follows, using the analytical model in Section \ref{sec:3}, 
we analyse the multipole power spectra measured from the best-fitting mock 
LRG catalogue, and examine how well we can correctly recover the 
input cosmological parameters of the $N$-body simulations. 
\begin{figure*}
\includegraphics[width=100mm]{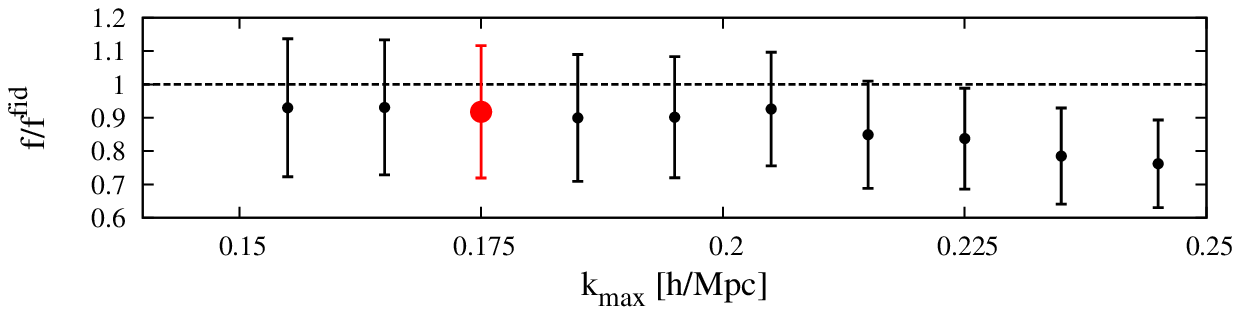}
\includegraphics[width=100mm]{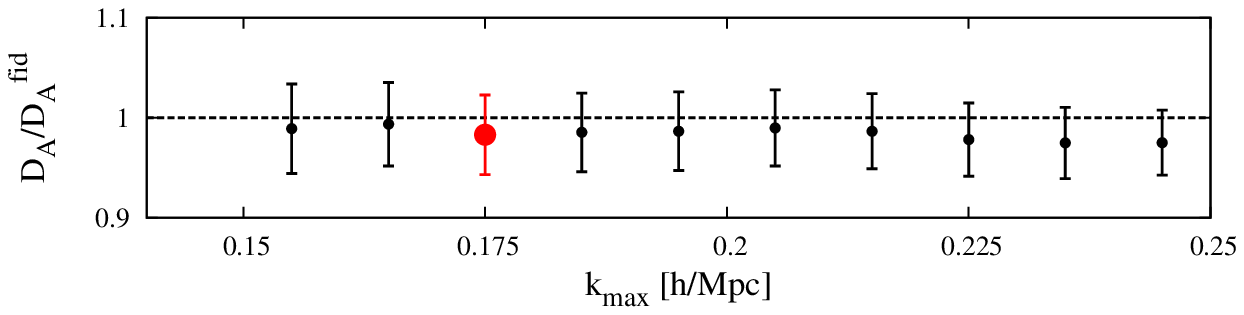}
\includegraphics[width=100mm]{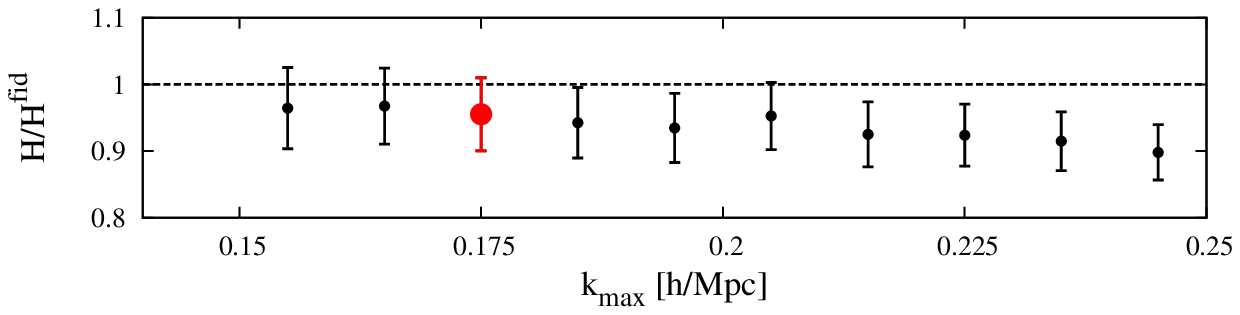}
\includegraphics[width=100mm]{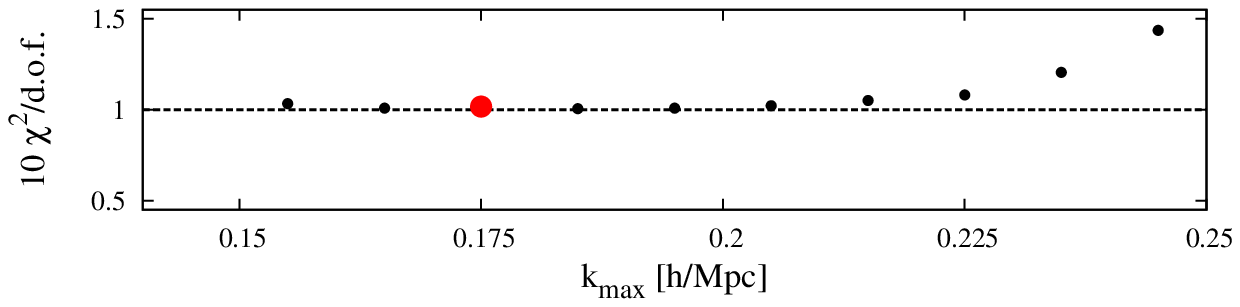}
\caption{Performance of our fiducial model against the mock catalogues. 
We plot the best-fitting parameters for {\bf WMAP5-z035} as a function of $k_{\rm max}$. 
The horizontal dotted lines show the fiducial values of the parameters. 
As is clear from this figure, the best-fitting parameters are 
consistent with the fiducial ones within 1-$\sigma$ error up to 
$k_{\rm max} = 0.175 [h$/Mpc], which is marked as the large (red; colors are available for the online version) circle.  
For $k_{\rm max} > 0.175 [h$/Mpc], the perturbative approach 
breaks down because of the non-linearity.
We plot
$10\chi^2$/d.o.f. in the bottom panel because we adopt the error on the power spectrum 
estimated for the SDSS DR7 LRG sample while the analysed mock spectra are 
measured from the total volume 11 times larger than the observation.}
\label{fig:1D_params}
\end{figure*}

\subsubsection{Method}

We perform the MCMC analysis and try to fit the mock power spectrum 
with our analytical model under different assumptions. 
We investigate the sensitivity of the results to the maximum wavenumber ($k_{\rm max}$) included in the analyses, 
prior cosmological assumptions and the effective redshift of the galaxy sample, 
different theoretical models for RSDs and galaxy bias.
We examine six different setups labeled as follows:
\begin{itemize}
\item {\bf WMAP5-z035}: 
We here employ the full PT model, i.e., TNS model for RSDs with the linear scale-dependent galaxy bias given in equations~(\ref{eq:TNS_linear_bias}) and (\ref{eq:powerlaw}), taking a proper account of the AP effect 
[equations(\ref{eq:AP})-(\ref{alpha})]. In computing the model power spectra,  
we adopt the same cosmological parameters as those used in the $N$-body simulations except for $D_{\rm A}$, $H$ and $f$,
and evaluate the spectra at $z = 0.35$, corresponding to the output redshift of the simulations. 

\item {\bf WMAP5-P0P2}: Same as {\bf WMAP5-z035}, 
but in fitting the model to the mock catalogue, we use only $P_0$ and $P_2$.

\item {\bf WMAP5-z03}: Same as {\bf WMAP5-z035}, but the model power spectra are evaluated at $z = 0.3$, 
slightly different from the output redshift $z=0.35$.

It is indeed nontrivial to know an effective redshift of the DR7 LRG sample, 
and there is also an ambiguity in the effective redshift. Further, an evolution effect of LRGs 
over a wide redshift range, known as the light-cone effect \citep{Yamamoto:1999hc}, 
might lead to an misinterpretation of the cosmological results. We thus 
check if theoretical template at a slightly different redshift can correctly recover the cosmological parameters.

\item {\bf Planck}: Same as {\bf WMAP5-z035}, but in computing the model power spectra, 
we adopt the cosmological parameters suggested by the Planck observation 
\citep{Planck-Collaboration:2013fr}; 
$\Omega_{\rm m} = 0.32$, $\Omega_{\rm b} = 0.0496$, $h = 0.67$, $n_s = 0.96$, $\sigma_8 = 0.809$. 

\item {\bf WMAP5-noAB}: Same as {\bf WMAP5-z035}, but in computing the model power spectra, 
the $A$ and $B$ terms in the TNS model (\ref{eq:TNS}) are 
dropped out. A comparison with {\bf WMAP5-z035} will show how much 
details of the modeling of RSD can affect the result of cosmological 
parameter estimations. 

\item {\bf WMAP5-cbias}: Same as {\bf WMAP5-z035}, but the galaxy bias model 
(\ref{eq:bias}) is replaced with a constant bias, i.e. $b(k) = b_0$.
\end{itemize}

The setup of the MCMC analysis is summarized as follows. 
The total number of free parameters in our MCMC analysis is 7; 
the linear growth rate ($f$), the angular diameter distance ($D_{\rm A}$), the Hubble parameter ($H$), 
the one-dimensional velocity dispersion in equation~(\ref{eq:TNS}) ($\sigma_v$), 
and the bias parameters in equation(\ref{eq:bias}) ($b_0, A_1, A_2$), 
except for {\bf WMAP5-cbias}.

We find the best-fitting parameter set in the seven-dimensional parameter space 
by minimizing the chi-squared defined by 
\begin{align}
\label{eq:chi2_MCMC} 
\chi^2=\sum_{\ell=0,2,4}\sum_{i=1}^{N_{{\rm bin}}}\Big[\frac{P_{\ell}^{\rm model}(k_i)-P_{\ell}^{\rm obs}(k_i)}{\Delta P^{{\rm obs}}_{\ell}(k_i)}\Big]^2,
\end{align}
where $P_{\ell}^{\rm model}(k_i)$ is defined by equation~(\ref{eq: P_ell^model}) 
and $P_{\ell}^{\rm obs}(k_i)$ is the measured multipole power spectra. 
Note that $\ell=4$ is not included in the case of {\bf WMAP5-P0P2}. 
The error, $\Delta P_{\ell}^{\rm obs}(k_i)$, is given by equation~(\ref{eq:error}). 
Since we want to figure out the significance of possible systematics relative to the real observational errors, 
we will adopt the statistical error of the DR7 LRG to estimate 
$\Delta P_{\ell}^{\rm obs}(k_i)$. In equation~(\ref{eq:chi2_MCMC}), 
the quantity $N_{\rm bin}$ denotes the number of $k$-bins, which depends on 
the maximum wavenumber $k_{\rm max}$ used for the parameter estimation. 
Below, we demonstrate how the MCMC results depends on 
$k_{\rm max}$, and determine the appropriate value in our analysis.
In the MCMC analysis, we use a part of publicly-available MCMC code {\tt CosmoMC} 
\citep{Lewis:2002qf}.

\begin{figure}
\includegraphics[width=92mm]{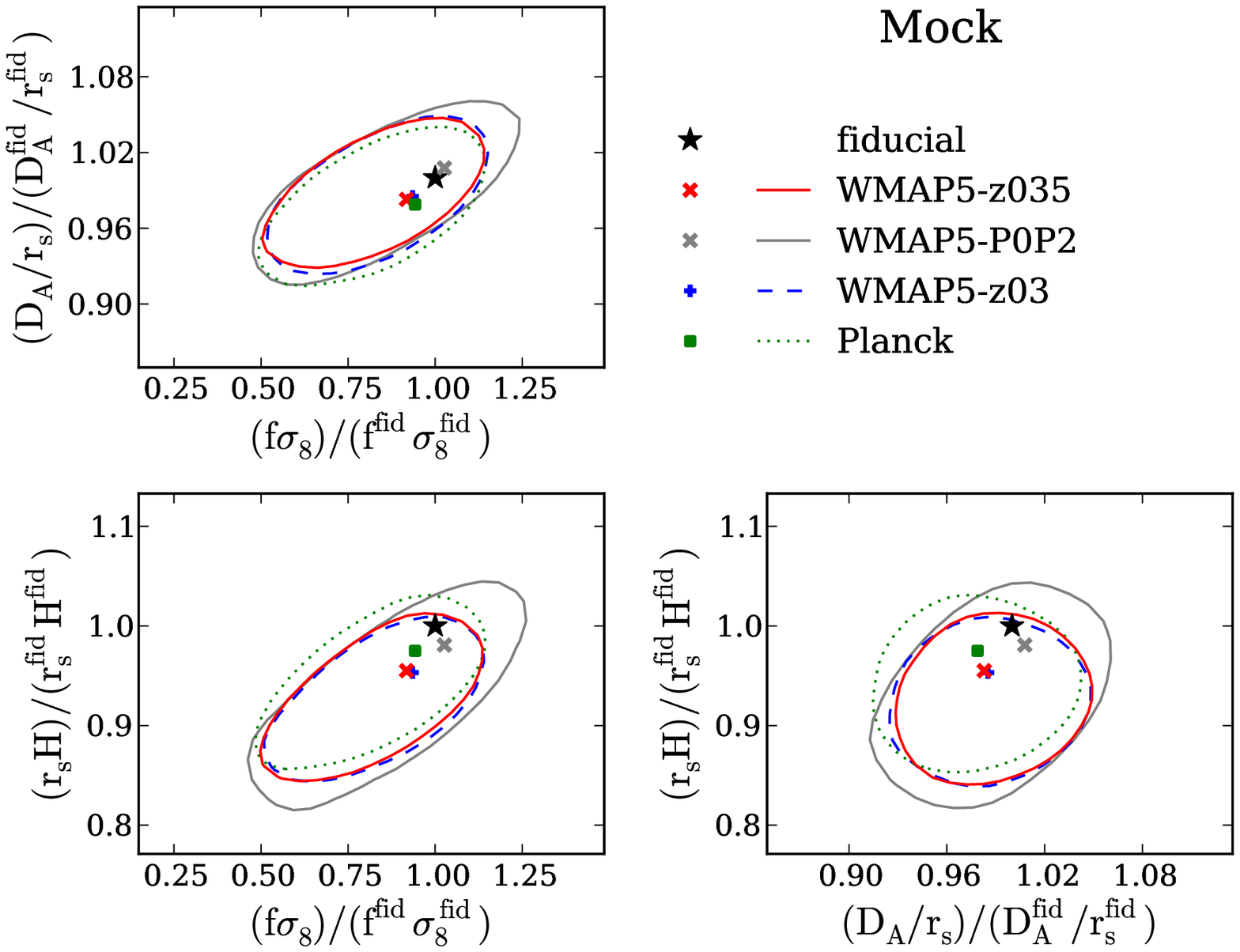}
\includegraphics[width=92mm]{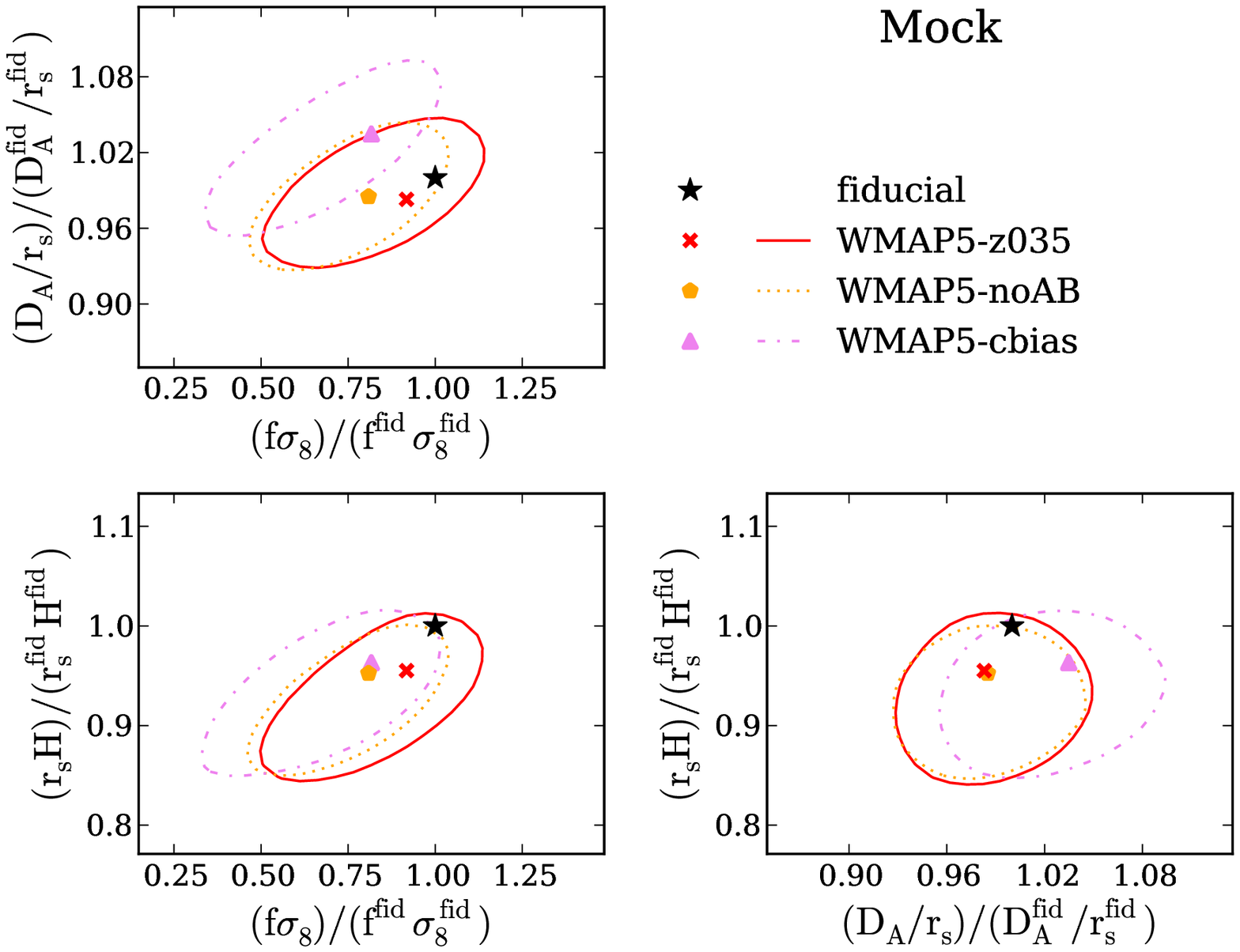}
\caption{
Test of systematics of our analysis using the mock catalogues
for estimating the linear growth rate, angular diameter distance, 
and Hubble parameter. 
The star in each panel is the fiducial input parameters. 
We plot the best-fitting results with symbols and 68 \% confidence contours for the different setup of the analysis  (symbols and contours are plotted with the same color; 
available for the online version) , {\bf WMAP5-z035},
{\bf WMAP5-P0P2}, {\bf WMAP5-z03},
{\bf Planck}, {\bf WMAP5-noAB}, {\bf WMAP5-cbias}, as noted in 
the figure. 
This figure shows that our model correctly recovers the fiducial 
cosmological model, and the incorrect cosmological assumptions 
are only marginal when taking the statistical error similar to
the observed power spectra of LRGs in the SDSS DR7 into account. 
Note that in upper three panels, the distance scales $D_{\rm A}$ and $H$ 
are also normalized by the sound horizon scales at baryon drag epoch, $r_s$, to highlight 
a difference in the measurement of distance scales themselves.
 Also, the linear growth rate is scaled by $\sigma_8$. 
In lower three panels, the ratio $r_s/r_{s, \rm fid}$ and $\sigma_8/\sigma_{8, \rm fid}$ are unity because all the theoretical template is computed with the same underlying cosmology.
}
\label{fig:mock_2D}
\end{figure}

\begin{figure}
\includegraphics[width=80mm]{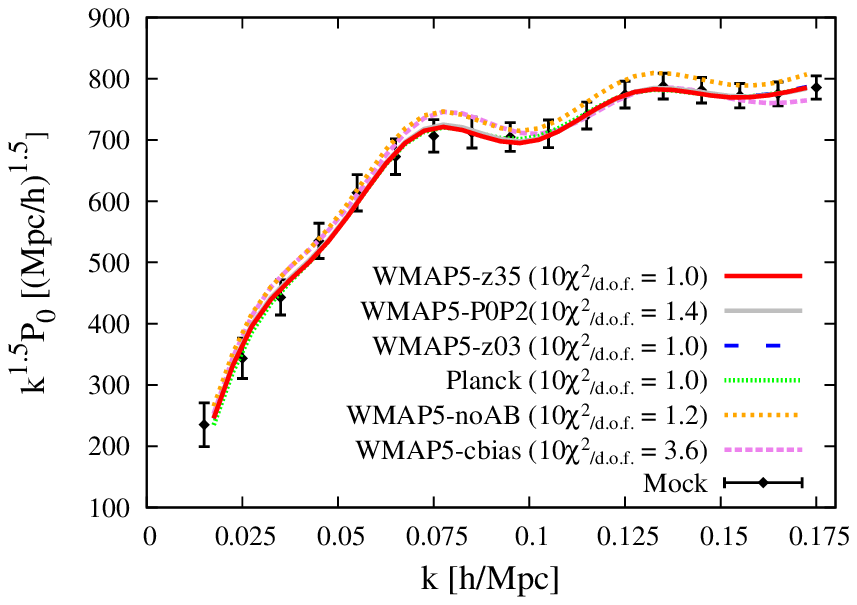}
\includegraphics[width=80mm]{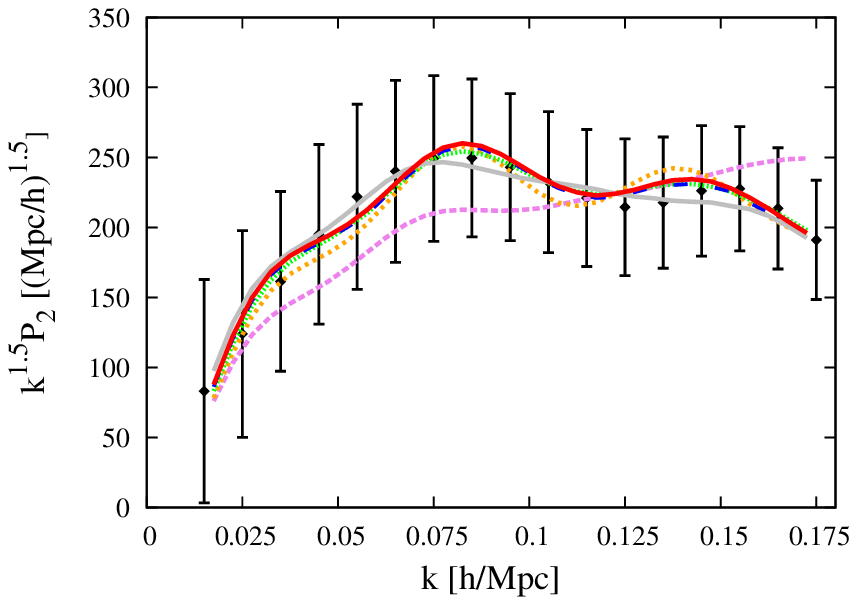}
\includegraphics[width=80mm]{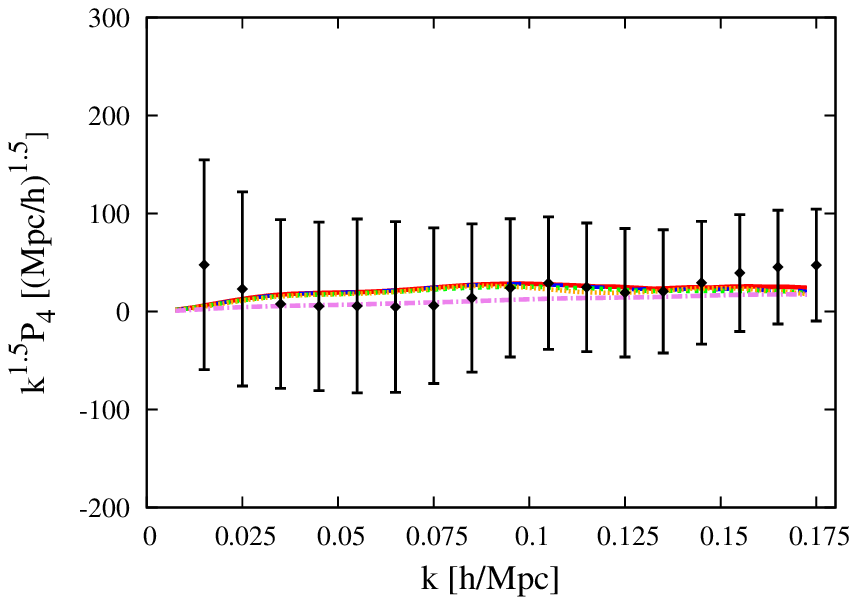}
\caption{
Comparisons of the multipole spectrum from the mock 
catalogues and the best-fitted curves for the different 
setups in Fig.~\ref{fig:mock_2D}.
The meaning of the curves is noted in the top panel, 
the same as those of  Fig.~\ref{fig:mock_2D}. 
The diamond shows the multipole power spectra measured 
from our mock catalogue. Note that the error bars show the 
statistical error for the SDSS DR7 LRG sample, and do not reflect the uncertainties 
of the mock power spectrum.
}
\label{fig:fitmockpower}
\end{figure}

\subsubsection{Results}

In order to determine the applicable range of our theoretical template, 
we first examine the {\bf WMAP5-z035} case, and perform the MCMC analysis 
varying the maximum wavenumber $k_{\rm max}$ from 0.155 to 0.245 [$h$/Mpc].
Figure~\ref{fig:1D_params} shows the goodness of fitting (i.e. 10$\chi^2$/d.o.f.) and the marginalized $1$-$\sigma$ confidence
intervals of the 3 parameters, $f$, $D_{\rm A}$, and $H$ as a function of $k_{\rm max}$.
As is seen from this figure, the best-fitting values correctly 
reproduce the fiducial value well within the 1-$\sigma$ statistical confidence for 
$k_{\rm max} \leq 0.175$ [$h$/Mpc]. 
The results look quite reasonable in the sense that 
the TNS model is accurate at a few percent level for the 
matter power spectrum at $z=0.35$ \citep{Taruya:2012ai,Taruya:2013qy}.
At $k> 0.175\,[h$/Mpc], the model prediction is known to deviate from dark matter simulations. 
Even though the estimated values of the 3 parameters are barely within 
1-$\sigma$ statistical confidence at $k_{\rm max} = 0.205 [h/$Mpc],
the value of $H/H_{\rm fid}$ is not properly estimated  
at $k_{\rm max} \geq 0.185$ [$h$/Mpc]. 
This behavior presumably comes from some flaw of our analytical model. 
Thus, we conservatively adopt $k_{\rm max} = 0.175 [h/$Mpc] in the following 
MCMC analysis, which corresponds to the number of bins, $N_{\rm bin} = 17$.

We notice that the estimated values of $(f, D_A, H)$ are a bit systematically lower than the fiducial values at any $k_{\rm max}$.
Since the underestimation is regarded as a systematic error in our modeling,
these will be taken into account in the final result (see Section \ref{sec:5}) as a systematic error to the total error budget.
The systematic errors are evaluated as ($\pm8.2\%, \pm1.8\%, \pm4.5\%$) for $(f, D_A, H)$, respectively, at $k_{\rm max} = 0.175 [h/$Mpc].

Let us now investigate possible systematics due to
an incorrect cosmological prior or model assumptions. 
The results are summarized in Figs.~\ref{fig:mock_2D} and \ref{fig:fitmockpower}. 
In Fig. \ref{fig:mock_2D}, each pair of the symbol and curve (with the same color; 
available for the online version) shows the best-fitting value and the 68 \% confidence 
contour from our fitting with the different setups, 
{\bf WMAP5-z035} (solid red curve), 
{\bf WMAP5-P0P2} (solid gray curve),
{\bf WMAP5-z03} (dashed blue curve),
{\bf Planck} (dotted green curve), 
{\bf WMAP5-noAB} (short-dotted orange curve),
{\bf WMAP5-cbias} (dot-dashed pink curve). 
The star in each panel is the fiducial input parameters. 
The upper three panels in Fig.~\ref{fig:mock_2D} show that the input 
parameters are recovered even if the prior assumption is slightly incorrect.  
The results of {\bf WMAP5-z035} and {\bf WMAP5-z03} are almost identical, which means 
that a choice of the effective redshift is not important. 
The contour of {\bf WMAP5-z035} agrees with {\bf Planck}, which means that the difference 
of cosmological parameters between {\bf WMAP5-z035} and {\bf Planck} does not
systematically bias the results. 

In these panels, we plot the combination $(f \sigma_8)/(f^{\rm fid}\sigma_{8}^{\rm fid})$, $[(D_{\rm A}/r_{\rm s})/(D_{\rm A}^{\rm fid}/r_{\rm s}^{\rm fid})]$, and $(H r_{\rm s})/(H^{\rm fid}r_{\rm s}^{\rm fid})$ where $r_s$ is the sound horizon scale at the baryon drag epoch.
Thus, Fig.\ref{fig:mock_2D} highlights a difference in the measurements of the linear growth rate and distance scales themselves.

Here the sound horizon scale $r_{\rm s}$ is numerically evaluated with {\tt CAMB} \citep{Lewis:2000ul}. 
Also, by comparing {\bf WMAP5-P0P2} with {\bf WMAP5-z035}, one sees that the hexadecapole 
improves the constraint marginally, which is qualitatively consistent with the Fisher matrix 
analysis \citep{Taruya:2011kx}. 
In lower three panels in Fig.~\ref{fig:mock_2D}, we demonstrate that our modeling of 
RSDs and galaxy bias successfully works in the sense that the best-fitting values of 
{\bf WMAP5-z035} are closest to the fiducial input values compared to the other cases,
{\bf WMAP5-noAB} and {\bf WMAP5-cbias}. 
However, the 1-$\sigma$ contours of {\bf WMAP5-noAB} 
contain the input parameters.
These two setups, {\bf WMAP5-noAB} and {\bf WMAP5-cbias}, are also worse than {\bf WMAP5-z035} in terms of the goodness of fitting. The value of $10 \chi^2/$d.o.f. for {\bf WMAP5-noAB} and {\bf WMAP5-cbias} is respectively 1.2 and 3.6 (see the legends in Fig.~\ref{fig:fitmockpower}). We compare the best-fitting curves of the different setups in Fig. \ref{fig:fitmockpower}.

The authors of \citet{Taruya:2010lr,Nishimichi:2011xi,Ishikawa:2013yq} adopted a similar 
modeling of RSDs and galaxy bias, and showed a successful performance for the distribution  
of dark matter or haloes in the $N$-body simulations. 
Our results show that our modeling works for the galaxy mock catalogue as well.

\section{COSMOLOGICAL ANALYSIS WITH THE SDSS DR7 LRG catalogue}
\label{sec:5}

Let us now consider a simultaneous constraint on the cosmological parameters
from the multipole power spectra of the DR7 LRGs, applying the method examined 
in the previous section. 
We compute the model power spectra, assuming a flat $\Lambda$CDM model with 
$(\Omega_{\rm m}, \Omega_{\rm b}, h,n_s,\sigma_{8})=(0.32,0.0496,0.67,0.96,0.809)$ 
favored by the Planck result \citep{Planck-Collaboration:2013fr}. 
Note that we assume the same redshift-distance relation 
as that in measuring the multipole power spectra from the DR7 LRG catalogue.\par 
In the analysis of this section, we use the multipole power spectra of the DR7 
LRGs up to $\ell=4$ within the range of the wavenumber $k\leq k_{\rm max}=0.175\,[h$/Mpc] 
(see Section \ref{sec:2}), which includes $N_{\rm bin}=17$ equally-spaced bins for each 
multipole, unless explicitly stated otherwise.

\subsection{Simultaneous constraints on $f$,~$D_{\rm A}$, and $H$}
\label{sec:5.1}

\begin{figure}
\includegraphics[width=90mm]{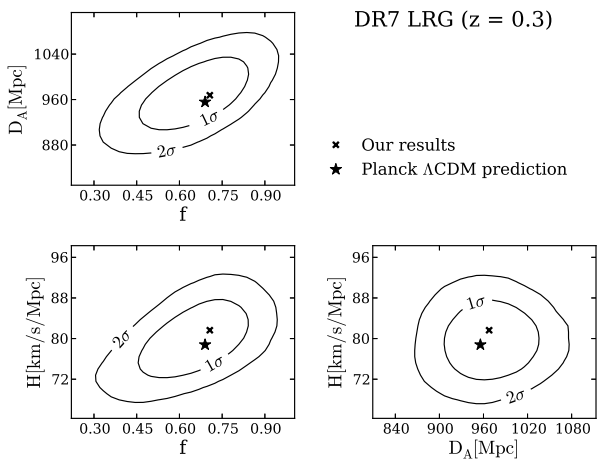}
\caption{
Simultaneous constraints on $f$, $D_{A}$, and $H$ from the multipole power 
spectra of the SDSS DR7 LRG sample.
In each panel, the inner and outer contours respectively represent 68 and 95 \% confidence levels.
We plot the best-fitting results in cross symbols as well as the $\Lambda$CDM prediction with the Planck cosmological parameters in stars.
}
\label{fig:SDSS_2D}
\end{figure}

\begin{center}
\begin{table*}
\scalebox{0.8}{
\begin{tabular}{c c c c c c c c }
\hline \hline
Setup &  Redshift & $\chi^2$/d.o.f. & $f (z)$ & $D_{\rm A} (z)$ [Mpc] & $H (z)$ [km/s/Mpc] \\
\hline \hline
{\bf Planck-z03} (canonical) & 0.3 & 0.45 & 0.71 $\pm_{stat.}$ 0.12$\pm_{sys.}0.06$ & 968 $\pm_{stat.}$ 42$\pm_{sys.}17$  & 81.7 $\pm_{stat.}$ 5.0$\pm_{stat.}3.7$ \\
\hline
{\bf Planck-z035} & 0.35 & 0.50 & 0.70 $\pm_{stat.}$ 0.13$\pm_{sys.}0.06$ & 961 $\pm_{stat.}$ 40$\pm_{stat.}0.17$ & 78.8 $\pm_{stat.}$ 4.4$\pm_{sys.}3.5$  \\ \hline
{\bf WMAP5} & 0.3 & 0.46 & 0.71 $\pm_{stat.}$ 0.13$\pm_{sys.}0.06$ & 1012 $\pm_{stat.}$ 41$\pm_{sys.}18$ & 78.3 $\pm_{stat.}$ 4.4$\pm_{sys.}3.5$ \\ \hline
{\bf Planck-noAB} & 0.3 & 0.48 & 0.65 $\pm_{stat.}$ 0.12$\pm_{sys.}0.05$ & 967$\pm_{stat.}$39$\pm_{sys.}17$ & 80.7 $\pm_{stat.}$ 3.9$\pm_{sys.}3.6$ \\ \hline
{\bf Planck-cbias} & 0.3 &0.68 & 0.68 $\pm_{stat.}$0.13$\pm_{sys.}0.06$ & 1041$\pm_{stat.}$41$\pm_{sys.}19$ & 81.6$\pm_{stat.}$5.0$\pm_{sys.}3.7$  \\ \hline
\end{tabular}}
\caption{
Test of systematics with the SDSS DR7 LRG sample, adopting the similar setups as those in Section \ref{sec:4.3}. 
The best-fitting parameters and the goodness of fitting ($\chi^{2}$/d.o.f.) are listed. We adopt {\bf Planck-z03} 
as a canonical setup, and the final results are obtained with this setup.
 The error denoted by $sys.$ is the possible systematic error evaluated in the mock catalogue (see Section \ref{sec:4.3}). 
}
\label{tab:systematic}
\end{table*}
\end{center}

Our best-fitting model is shown together with the DR7 LRG $P_{\ell}(k)$ in Fig.~\ref{fig:fitSDSSpower}. 
Simultaneous constraints on $f$, $D_{\rm A}$, and $H$ marginalized over the other model parameters are 
presented in Fig. \ref{fig:SDSS_2D}. 

The value of $\chi^2$/d.o.f for our best-fitting model is 0.45,
which is somewhat smaller than the expectation ($\chi^2$/d.o.f. $\sim$ 1).
The reason for the small $\chi^2$/d.o.f. may partly come from the fact that we neglect the covariance both between different $k$s and $\ell$s.
We expect that full treatment of the covariance matrix can increase the $\chi^2$
although we do not fully understand the reason why $\chi^2$ is small.
The estimated parameters, however, will not change significantly even when
the off-diagonal components of covariance are taken into account as discussed in\citep{Takahashi:2009kl,Takahashi:2011lr}.

In Fig.~\ref{fig:SDSS_2D}, we compare our results with the values predicted by the Planck best-fitting $\Lambda$CDM cosmology and find no evidence of significant discrepancy.

Let us discuss the degeneracy between $(f,D_{\rm A},H)$ and the other nuisance 
parameters. The FoG parameter, $\sigma_{\rm v}$, is strongly degenerated 
with the linear growth rate, $f$ (the correlation coefficient $r(f,\sigma_{\rm v})=-0.62$). 
Also, the Hubble parameter, $H$, is moderately correlated with $f$ and $\sigma_{\rm v}$ 
($r(H,f)=0.53$ and $r(H,\sigma_{\rm v})=-0.74$, respectively). 
These facts are not surprising since $f$, $\sigma_{\rm v}$, and $H$ are all sensitive to the higher 
multipoles ($\ell=2$ and $4$), where a proper modeling of nonlinearity RSDs is essential. 
There is no significant degeneracy with the bias parameters, $A_1$ and $A_2$, 
although the linear bias parameter, $b_{0}$, has non-negligible correlations with 
$D_{\rm A}$ and $H$ ($r(b_{0}, D_{A})=0.40$ and $r(b_{0},H)=-0.75$, respectively). 

Nonetheless, these degeneracies are not perfect. This fact implies that the power spectrum amplitude adds information on the geometric parameters, as opposed to the isotropic case. This is explained as follows. In principle, through the A-P effect, the power spectrum amplitude depends not only on $b_0^2$ but also on $H/D_A^2$. But the degeneracy between these parameters cannot be broken without other extra information. With the measurement of BAO scale and RSD 
(roughly speaking, the quadrupole-to-monopole ratio, $P_0/P_2$), however,  
we can simultaneously estimate $D_{A}^{2}/H$ and $f\sigma_8$, free from the bias parameters. Then, if we add the amplitude and shape information in the anisotropic power spectrum, which respectively depends on 
$b_{0}^{2}H/D_{A}^{2}$ and $b_0^2D_{A} H$, the degeneracy between $b_0$, $D_A$ and $H$ is broken, and we can separately determine the geometric parameters \citep{Padmanabhan:2008xc,Anderson:2012ap,Percival:2010dq,Percival:2007cy}.


We see that the correlations between the linear bias parameter, $b_{0}$, 
and the geometric parameters, $D_A$ and $H$ are not perfect. 
This implies that the power spectrum amplitude may add information on 
the geometric parameters. As opposed to the isotropic case, this makes
when the anisotropic part of the power spectrum is included 
in the following reason. The BAO scales and RSD 
(roughly speaking, the quadrupole-to-monopole ratio, $P_0/P_2$) 
respectively well constrain $D_{A}^{2}/H$ and $f\sigma_8$ free from bias parameters.
The degeneracy between $D_A$ and $H$ is broken by both amplitude and shape information in the anisotropic power spectrum, which respectively depends on 
$b_{0}^{2}H/D_{A}^{2}$ and $b_0^2D_{A} H$ \citep{Padmanabhan:2008xc,Anderson:2012ap,Percival:2010dq,Percival:2007cy}.

\par 

Although we have already presented our main results, it would be still worthwhile mentioning
how robust our modeling is against different setups as a check. Here we go 
through a similar study to what we have done for the mock catalogue. 
Namely, we compare constraints using several slightly different setups which are summarized 
in Table.~\ref{tab:systematic}.
The labels of the setup in Table.~\ref{tab:systematic} are summarized as follows.

\begin{itemize}
\item {\bf Planck-z03}:  The fiducial model described in the above. 
The redshift at which we evaluate the model power spectra is $z = 0.3$.
\item {\bf Planck-z035}: Same cosmological model as {\bf Planck-z03} but we evaluate the model 
power spectra at $z = 0.35$, which is normally quoted as the effective redshift in the 
FKP-type measurement \citep{Percival:2010dq}.
\item {\bf WMAP5}: We assume the cosmological parameters favored by the WMAP 5-yr result, 
$\Omega_{\rm m} = 0.28$, $\Omega_{\rm b} = 0.046$, $h = 0.7$, $n_s = 0.96$, $\sigma_8 = 0.8$,
and  $z = 0.3$, for computing the model spectra. 
\item {\bf Planck-noAB}: Same as {\bf Planck-z03}, but with the A and B correction terms 
in the RSD model (\ref{eq:TNS}) dropped out.
\item {\bf Planck-cbias}: Same as {\bf Planck-z03}, but with the galaxy bias (\ref{eq:bias}) 
being a constant, i.e., $b(k) = b_0$.
\end{itemize}

As is seen from Table.~\ref{tab:systematic}, {\bf Planck-z03} gives the smallest $\chi^{2}$, while 
the difference is small. The constraints on $(f,D_{\rm A},H)$ are all consistent 
with each other, excepting {\bf Planck-cbias}. 
The bias parameters are in fact more important than the others in order to well fit to 
the monopole. Comparison between {\bf WMAP5} and {\bf Planck-z03} shows that our constraints are not 
sensitive to choice of the underlying cosmology for the model power spectrum. 
We thus conclude that our results are robust against such systematics.

\begin{figure}
\includegraphics[width=80mm]{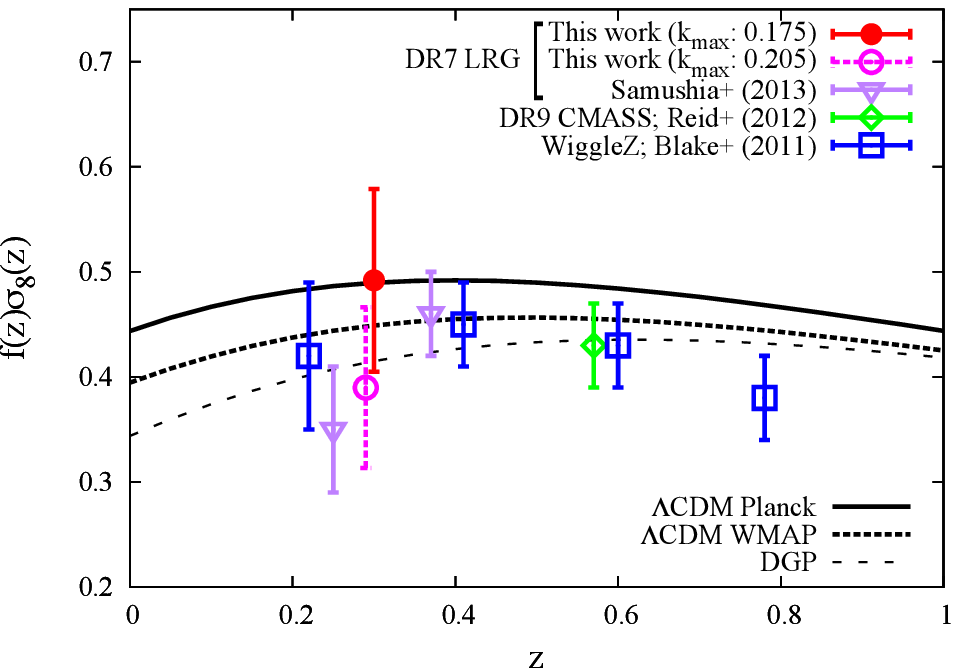}
\includegraphics[width=80mm]{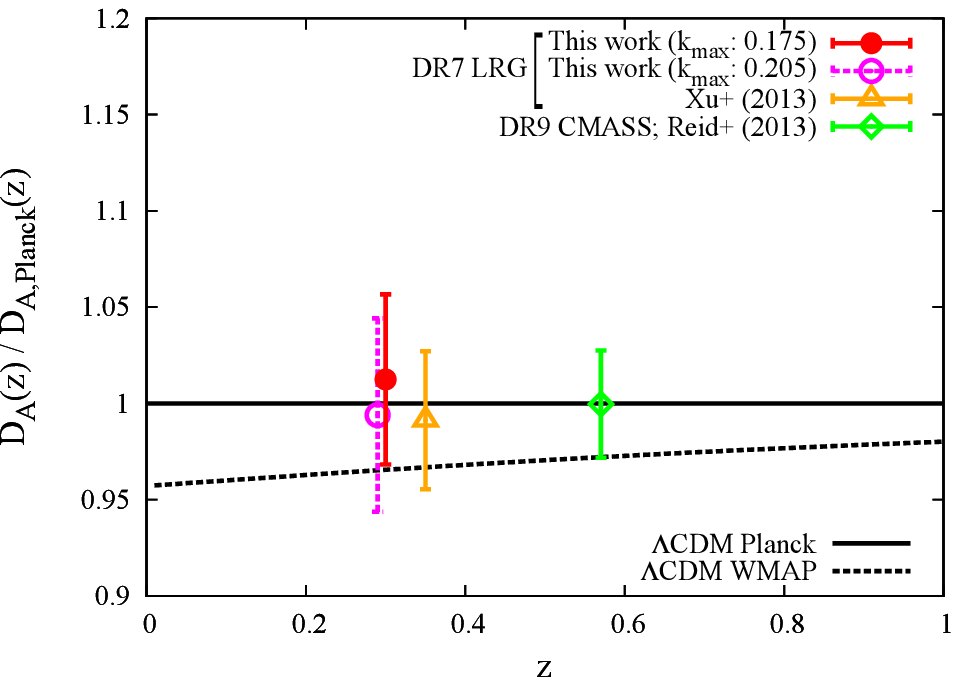}
\includegraphics[width=80mm]{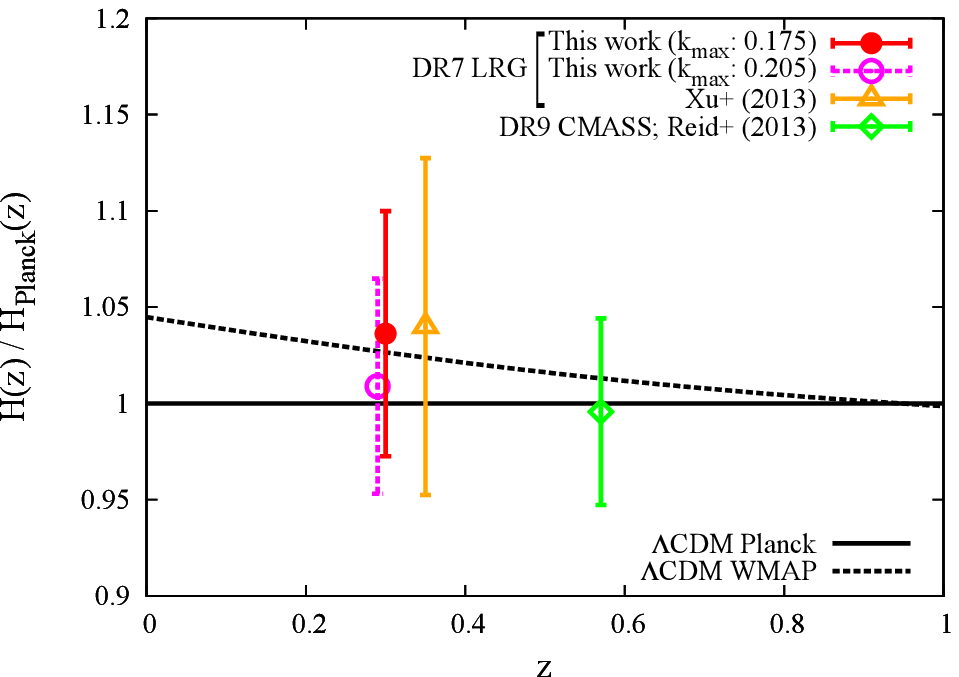}
\caption{
Comparisons of our results with those of previous works and 
model predictions.  
The linear growth rate (top), angular diameter distance (middle), 
and Hubble parameter (bottom) as a function of redshift are shown. 
We plot our results with $k_{\rm max} = 0.175 [h/$Mpc] in filled (red; colors are available for the online version) circles as well as our aggressive results with $k_{\rm max} = 0.205 [h/$Mpc] in open (magenta) circles to caveat the systematic due to non-linear RSDs.
For comparison, we also display an open (purple) inverted triangle
from \citep{Samushia:2012ce},
open (blue) boxes from \citep{Blake:2011fj}, 
open (green) diamonds from \citep{Reid:2012ly}, 
and open (orange) triangles from \citep{Xu:2013fr}. 
 The solid curve is the prediction of the flat $\Lambda$CDM assumption with the Planck cosmological parameters 
($\Omega_{\rm m} = 0.32$, $h = 0.67$) \citep{Planck-Collaboration:2013fr} 
and the dotted curve is those of the WMAP cosmological parameters 
($\Omega_{\rm m} = 0.279$, $h = 0.701$) \citep{Komatsu:2009ly}. 
On the other hand, the dashed curve is the prediction of the DGP model \citep{Dvali:2000yq}.
 Note that we here do not include the systematic errors for our result. 
}
\label{fig:previous}
\end{figure}
\subsection{Comparison with previous works}
\label{sec:5.2}

Here let us mention the consistency of our results compared with previous works. 
We show some examples of similar works 
\citep{Blake:2011fj,Reid:2012ly,Samushia:2012ce,Xu:2013fr} in Fig.~\ref{fig:previous} 
and Table.~\ref{tab:previous}, together with the predictions from different cosmological models.
Fig.~\ref{fig:previous} shows that all of the results tend to 
underestimate $f\sigma_8$ compared to the Planck best-fitting $\Lambda$CDM model 
but no significant deviation from a $\Lambda$CDM model is confirmed. 
Our results are in a good agreement with those in \citet{Samushia:2012ce,Xu:2013fr}, in which 
the same galaxy sample, i.e., the DR7 LRGs is used with different statistics or setups.

The solid and dotted lines respectively show theoretical prediction in a flat $\Lambda$CDM model with the Planck and WMAP-5yr cosmological parameters.
We here also plot the linear growth rate of DGP model \citep{Dvali:2000yq} with Planck cosmological parameters, as one of the representative modified gravity models.

While we put such a simultaneous constraint using the multipole power spectra for the first time up 
to the hexadecapole moment, 
our results are consistent with previous works \citep{Samushia:2012ce,Xu:2013fr}. 
\citet{Samushia:2012ce} measured the linear growth rate, $f$ from the LRG sample,  
but they ignored the AP effect and used a different approach with the correlation function. 
On the other hand, \citet{Xu:2013fr} investigated the AP effect through the location of the BAO ring, 
marginalizing over the broadband shape information. Also, note that \citet{Xu:2013fr} adopted the 
reconstruction procedure of the BAO feature so that they can see the signal more clearly. 
Their measurement errors on $D_{\rm A}$ and $H$ 
(3.6\% for $D_{\rm A}$ and 8.4\% for $H$) 
are somewhat similar to what we obtain 
(4.3\% for $D_{\rm A}$ and 6.1\% for $H$), 
even though we utilize the broadband shape information in the anisotropic power spectrum. 
These results suggest that the impact of the AP effect on the isotropic part ($P_0$) is mostly constrained
through the shift of the location of the BAO signature, while the change of the overall amplitude and shape 
is somewhat absorbed in the bias function. On the other hand, the signature of BAOs on the anisotropic part 
($P_2$ and $P_4$) is not clear given the current level of the statistical error. Instead, the broadband shape of
these moments that can significantly be altered by the AP effect might give most of the information 
\citep{Padmanabhan:2008xc}, leading to the difference from the result in \citet{Xu:2013fr}.

Let us emphasize again that our study is the first attempt to constrain
simultaneously on the gravitational growth and the cosmic distance scale
especially with the multipole power spectra up to the hexadecapole ($\ell=4$).
\citet{Reid:2012ly} made a similar effort for the BOSS DR9 CMASS sample 
but they restrict the analysis to the monopole and quadrupole moments of the two-point correlation function. 
The two-point correlation function in principle carries the same cosmological information with the power spectrum 
but may suffer from somewhat different systematics issues \citep{Reid:2011kx}, and hence a consistency check 
between the two analyses would be important to validate the results.

\begin{figure*}
\includegraphics[width=100mm]{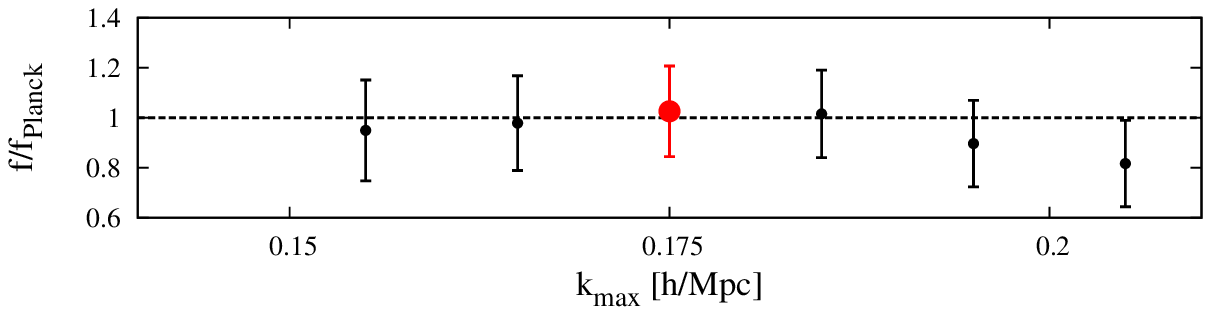}
\includegraphics[width=100mm]{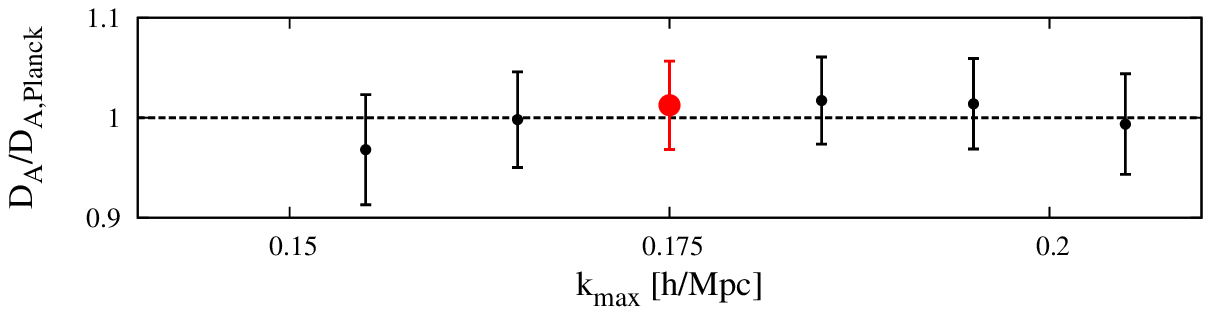}
\includegraphics[width=100mm]{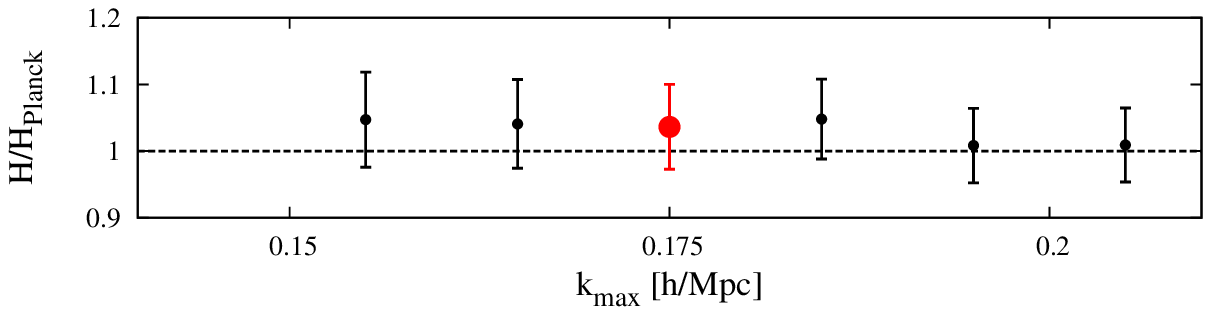}
\includegraphics[width=100mm]{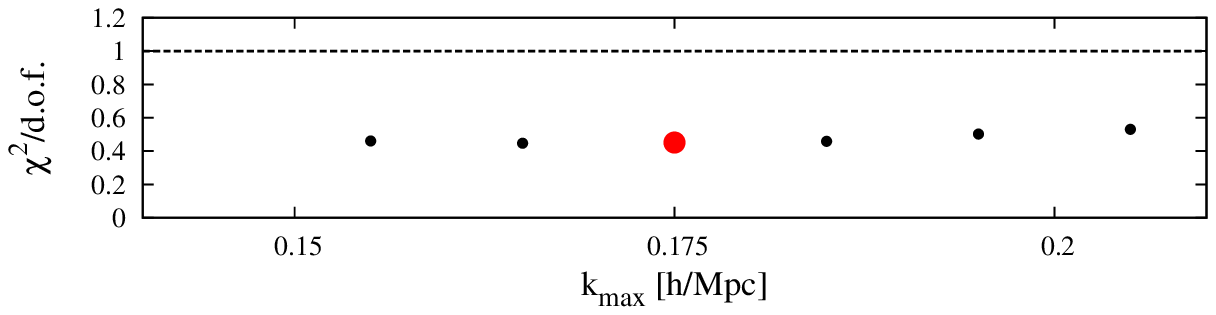}
\caption{
Similar figures to Fig.~\ref{fig:1D_params} but with the SDSS DR7 LRG sample. 
We plot the best-fitting parameters for {\bf Planck-z03} as a function of $k_{\rm max}$. 
The horizontal dotted lines show the $\Lambda$CDM prediction with 
the Planck cosmological parameters. 
The large (red; available for the online version) circle is
our canonical results with $k_{\rm max} = 0.175 [h$/Mpc].
A remarkable difference ($\sim$ 20\%) between the canonical results and the aggressive results 
with $k_{\rm max} = 0.205 [h$/Mpc] appears in estimated value of $f$, 
while such behavior was not seen in the analysis 
with the mock LRG catalogue (see Section \ref{sec:4.3}).
}
\label{fig:1D_params_SDSS}
\end{figure*}

\begin{figure}
\includegraphics[width=85mm]{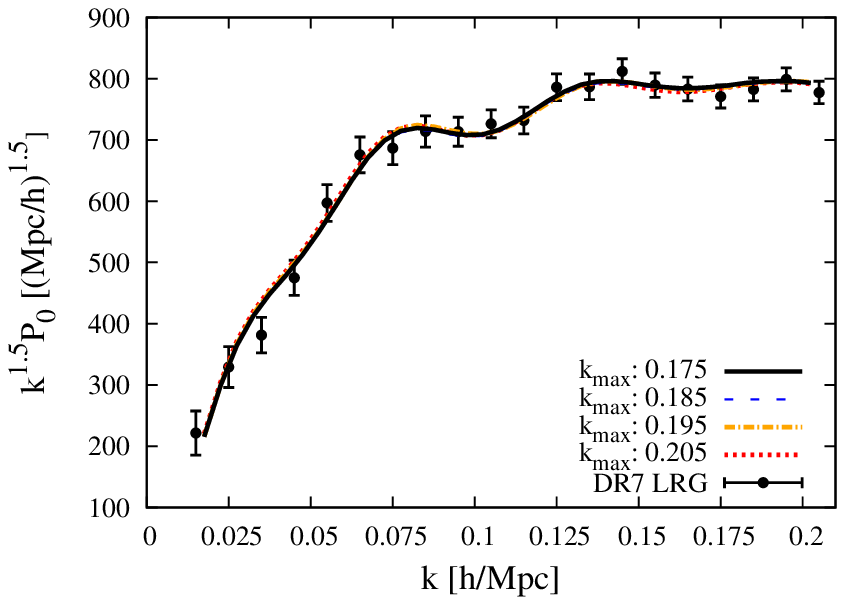}
\includegraphics[width=85mm]{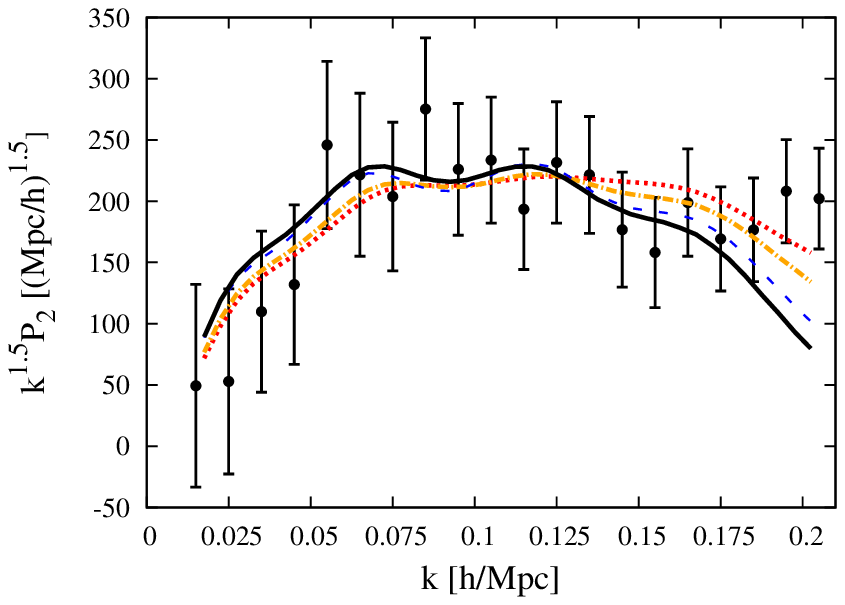}
\includegraphics[width=85mm]{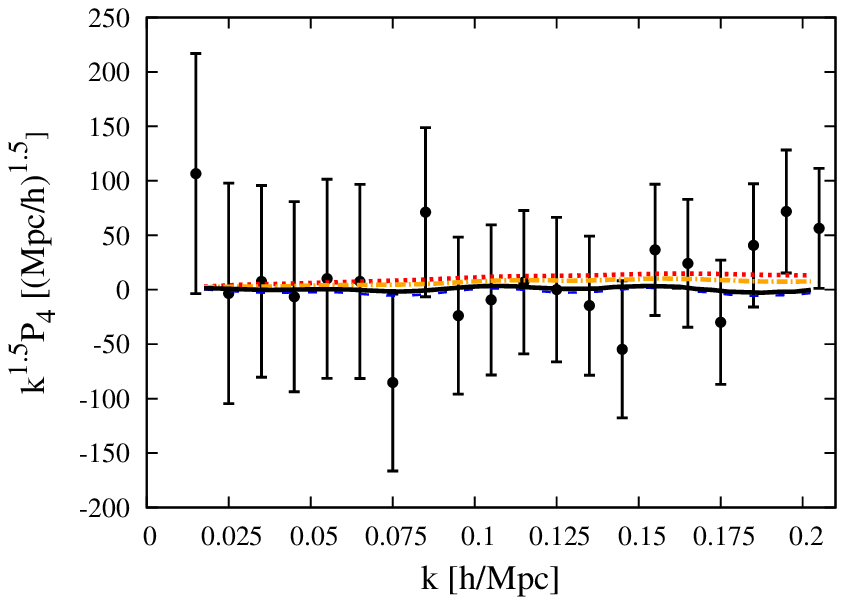}
\caption{
Comparisons of the best-fitting curves with the SDSS DR7 LRG sample with different $k_{\rm max}$ from 0.175 to 0.205 $[h/$Mpc].
We plot the best-fitting curves with $k_{\rm max} = 0.175 [h/$Mpc] in solid, $k_{\rm max} =0.185 [h/$Mpc] in (blue; colors are available for the online version) dashed , $k_{\rm max} = 0.195[h/$Mpc] in (orange) dot-dashed, 
and $k_{\rm max} = 0.205[h/$Mpc] in (red) dotted line. We also plot the measured multipole 
power spectra of the SDSS DR7 LRG sample in filled circles. 
}
\label{fig:k017-020}
\end{figure}

\subsection{What happens if aggressively fitted with higher $k_{\rm max}$?}
\label{sec:5.3}

Even though we have already presented the main results of this study, it might still be interesting to 
see what happens if we aggressively adopted a higher $k_{\rm max}$. One may wish to obtain
tighter constraints with adopting a higher $k_{\rm max}$. As we address in this paper, however, 
a smaller error does not necessarily assure a better constraint unless systematics both 
in the modeling and the measurements are well under control. 
In this subsection, we revisit a similar study to what we have done with the mock catalogues, 
and show how the results change as $k_{\rm max}$ is varied. 

Fig.\ref{fig:1D_params_SDSS} plots the one-dimensional constraints on $f$,$D_{\rm A}$, $H$, 
and the goodness of fitting, as a function of $k_{\rm max}$. 
Interestingly, there is a notable tension ($\sim 20\%$) between the derived values of 
the linear growth rate $f$ with $k_{\rm max} = 0.175$ and 
$0.205\,[h/$Mpc], while $D_{\rm A}$ and $H$ are in good agreement (Fig.\ref{fig:SDSS_2D}).
We have not observed such a behavior in the mock analyses (see Fig.~\ref{fig:1D_params}). 
In order to understand the cause of the discrepancy, we compare the best-fiting curves for 
each $k_{\rm max}$ in Fig.~\ref{fig:k017-020}. 
We argue that the discrepancy is driven by the fact that the measured quadrupole spectrum 
has data points somewhat larger than the line with $k_{\rm max} = 0.175\,[h/$Mpc]. 
This kind of feature in the quadrupole spectrum is not confirmed in the mocks, 
and it is hard to tell what really causes the behavior. 
One reason could be the sample variance which is suppressed by a large number of realizations 
of the mock catalogues. The limitation of our model is likely to be another reason. 
We have some signs from an analysis of the velocity statistics in the simulations that 
our treatment of the FoG suppression with a constant $\sigma_{v}$ does not fully capture 
the pairwise velocity statistics in the simulations (see e.g. \citet{Lam:2011aa} for a recent study on the pairwise velocity). 
Also, the effect of the 1-halo 
term could start to be dominant around the scales \citep{Valageas:2011tg,Hikage:2013zr}. 

\section{SUMMARY AND DISCUSSIONS}
\label{sec:6}
In this paper, we quantitatively study the anisotropic clustering of the SDSS DR7 LRG sample 
in order to simultaneously constrain the growth of structure via the RSDs 
and the cosmic distance scales via the AP effects. 
Using the multipole power spectra up to the hexadecapole ($\ell=4$), 
we obtain robust constraints on the linear growth rate $f(z = 0.3)=0.71\pm_{stat.}0.12\pm_{sys.}0.06$, 
the angular diameter distance $D_{\rm A}(z = 0.3)=968\pm_{stat.}42\pm_{sys.}17$[Mpc], and 
the Hubble parameter $H(z = 0.3)=81.7\pm_{stat.}5.0\pm_{sys.}3.7$[km/s/Mpc].
 Note that this result is based on $\sigma_8 (z=0.3) = 0.696$.

A remarkable point in this study is that we test our modeling 
systematics against `realistic' mock catalogues. Our mock catalogues consist 
of subhaloes identified in $N$-body simulations characterized by three parameters; 
the mass thresholds of host haloes ($M_{\rm min}^{\rm host}$) and subhaloes ($M_{\rm min}^{\rm sub}$), 
and satellite fraction $R_{\rm S}$. With a suitable choice of the parameter set (see Section~\ref{sec:4.2}), 
the subhalo catalogue quantitatively explain the clustering properties of LRGs \citep{Nishimichi:2013lr}, and it 
consistently reproduces the measured multipole power spectra. 
Then, we model the anisotropic galaxy power spectrum on the basis of 
perturbation theory. Combining with a phenomenological treatment 
of the galaxy bias, the robustness of our theoretical template is extensively tested against the subhalo catalogue. 
At a relevant redshift of the SDSS DR7 LRG sample, our model power spectra used as the fitting template are found 
to be valid up to $k_{\rm max}=0.175\,[h$/Mpc], 
and can correctly recover the input values of the underlying cosmological model. 
Hence, applying the same analytical model to the real observations, robust cosmological constraints 
on $f$, $D_{\rm A}$, and $H$ have finally been obtained.

The derived cosmological constraints are fully consistent with a flat $\Lambda$CDM cosmology.
The other cosmological results based on the anisotropic galaxy clustering are also 
consistent with a flat $\Lambda$CDM cosmology which 
implicitly assumes general relativity as the underlying theory of gravity. 
Although the measured values of the linear growth rate tend to slightly deviate from 
the Planck best-fitting $\Lambda$CDM model, more refined galaxy samples 
are definitely needed to statistically pin down the possible reasons of this. 
The galaxy samples with larger volumes, including the BOSS CMASS and LOWZ, continue to 
improve the measurement errors, and hence can be also used as more stringent tests of general relativity. 
With a sophisticated template taking account of the modification of gravity, 
we can further address the test of gravity beyond a consistency test, and put a 
tight constraint on theories of modified gravity \citep{Taruya:2013lr}.

As we have seen in Fig.~\ref{fig:k017-020} (see Section~\ref{sec:5.3}), 
an aggressive analysis of the power spectrum data up to a higher $k_{\rm max}$ 
results in a 20\% difference in the measurement of the linear growth rate. 
While this fact may be partly explained by the sample variance, 
it also implies that we do need a more elaborate modeling of non-linear RSDs 
if we want to push to smaller scales where higher signal-to-noise ratios are expected 
(see \citet{Hikage:2013zr} along this line). As increasing the statistical power, 
a more careful analysis combining the perturbation theory or a new theoretical 
framework with the simulations will be definitely important for robust cosmological constraints. 
We believe that the present approach with 
the subhalo catalogue provides a useful way to validate the RSD modeling, and can be 
generally applied to any galaxy redshift surveys. We hope to report such an analysis elsewhere.

\section*{ACKNOWLEDGEMENTS}
Authors thank Yasushi Suto for fruitful comments to the systematic study. 
Authors also thank Jun'ichi Yokoyama, Naoki Yoshida, Masahiro Takada, Takahiko Matsubara, Shirley Ho, and Chiaki Hikage for useful discussion.
AO acknowledges technical support from Toshiya Kashiwagi. 
This work is supported in part by a Grant-in-Aid from 
the Japan Society for the Promotion of Science (JSPS) 
(No.~24540257 for AT and No.~25887012 for SS). 
TN is supported by JSPS Postdoctoral Fellowships for Research Abroad. 
The research by K.Y. is supported in part
by Grant-in-Aid for Scientific researcher of Japanese Ministry of
Education, Culture, Sports, Science and Technology (No.~21540270 and
No.~21244033).

\bibliographystyle{mn2e}
\bibliography{oka_rsd}

\begin{landscape}
\begin{table}
\scalebox{0.87}{
\begin{tabular}{c c c c c c c c }
\hline \hline
Author & Sample & Measurement & Redshift & Galaxy bias & $f(z)\sigma_8(z)$ & $D_{\rm A}(z)$ [Mpc] & $H(z)$ [Mpc/km/s] \\
\hline \hline
This work & SDSS DR7 LRG & $P_0,P_2,P_4$ & 0.3 & $b_0(1+A_2k^2)/(1+A_1k)$ & 0.49 $\pm_{stat.}$ 0.08$\pm_{sys.}0.04$ ($f$ = 0.71$\pm_{stat.}$0.12$\pm_{sys.}0.06$) & 968$\pm_{stat.}$42$\pm_{sys.}17$ & 81.7$\pm_{stat.}5.0\pm_{sys.}3.7$ \\
\hline
\citet{Samushia:2012ce} & SDSS DR7 LRG & $\xi_0,\xi_2$ & 0.25 & constant (fix) & 0.35$\pm$0.06 & - & - \\
 &  & & 0.37 & constant (fix) & 0.46$\pm$0.04 & - & - \\ \hline
\citet{Xu:2013fr} & SDSS DR7 LRG & $\xi_0,\xi_2$ & 0.35 & constant (fix) & - & 1050$\pm$38 & 84.4$\pm$7.1 \\ \hline
\citet{} & SDSS DR7 LRG & $P(k,\mu) $ & 0.35 & constant (float) & - & 1050$\pm$38 & 84.4$\pm$7.1 \\ \hline
\citet{Reid:2012ly} & BOSS CMASS & $\xi_0,\xi_2$ & 0.57 & constant (float) & 0.43$\pm$0.069 & 2190$\pm$61 & 92.4$\pm$4.5 \\ \hline
\citet{Blake:2011fj} & WiggleZ & $P_0,P_2$ & 0.22 & constant (float) & 0.42$\pm$0.07 & - & - \\
 &  &  & 0.41 &  & 0.45$\pm$0.04 & - & - \\ 
 & & & 0.6 & & 0.43$\pm$0.04 & - & - \\ 
 &  & & 0.78 & & 0.38$\pm$0.04 & - & - \\ \hline
\citet{Contreras:2013ys} & WiggleZ & $\xi_0,\xi_2$ & 0.2 & constant (float) & 0.50$\pm$0.14 & - & - \\
 &  &  & 0.4 &  & 0.40$\pm$0.06 & - & - \\ 
 & & & 0.6 & & 0.37$\pm$0.08 & - & - \\ 
 &  & & 0.76 & & 0.42$\pm$0.09 & - & - \\ \hline
\end{tabular}}
\caption{
Comparison with previous works. Note that the normalization factor, $\sigma_8(z=0.3) = 0.696$, is fixed throughout this work.
Let us emphasize that we put the simultaneous constraints on $f$, $D_{\rm A}$, and $H$.
}
\label{tab:previous}
\end{table}
\end{landscape}

\end{document}